\documentclass[aps,prx,reprint,amsmath,amssymb,floatfix]{revtex4-2}
\usepackage{microtype}
\usepackage{graphicx}
\usepackage{float}
\usepackage{xcolor}
\usepackage{dcolumn}
\usepackage{multirow}
\usepackage{bm}
\usepackage[hidelinks,breaklinks,colorlinks,linkcolor=purple,citecolor=purple,urlcolor=purple]{hyperref}
\usepackage[capitalise,noabbrev]{cleveref}
\usepackage{tikz}
\usetikzlibrary{quantikz2}
\usepackage[linesnumbered,ruled,lined,noend,nosemicolon]{algorithm2e}
\usepackage[sort&compress]{natbib}
\creflabelformat{equation}{#2\textup{#1}#3}
\crefname{algocf}{Algorithm}{Algorithms}
\SetAlgoCaptionLayout{raggedright}

\SetCommentSty{mycommfont} 
\SetKw{And}{and}
\SetKw{Or}{or}
\SetKw{In}{in}
\DeclareMathOperator{\tr}{tr}
\DeclareMathOperator{\diag}{diag}
\DeclareMathOperator{\expt}{\mathbb{E}}
\DeclareMathOperator{\var}{\mathbb{V}}
\DeclareMathOperator{\cov}{Cov}
\DeclareMathOperator{\supp}{supp}
\DeclareMathOperator{\Pauli}{Pauli}
\renewcommand{\epsilon}{\varepsilon}
\newcommand{\tconsistent}{\diamond_T}
\begin{document}
\frenchspacing

\title{Scalable noise characterisation of syndrome extraction circuits\texorpdfstring{\\}{}
with averaged circuit eigenvalue sampling}
\author{Evan T. Hockings}
\email{evan.hockings@sydney.edu.au}
\author{Andrew C. Doherty}
\author{Robin Harper}
\affiliation{ARC Centre of Excellence for Engineered Quantum Systems,\\
School of Physics, The University of Sydney, Sydney, NSW 2006, Australia}
\date{April 1, 2025}

\begin{abstract}
Characterising the performance of noisy quantum circuits is central to the production of prototype quantum computers and can enable improved quantum error correction that exploits noise biases identified in a quantum device.
We develop a scalable noise characterisation protocol suited to characterising the syndrome extraction circuits of quantum error correcting codes, a key component of fault-tolerant architectures.
Our protocol builds upon averaged circuit eigenvalue sampling (ACES), a framework for noise characterisation experiments that simultaneously estimates the Pauli error probabilities of all gates in a Clifford circuit and captures averaged spatial correlations between gates implemented simultaneously in the layers of the circuit.
By rigorously analysing the performance of noise characterisation experiments in the ACES framework, we derive a figure of merit for their expected performance, allowing us to optimise their experimental design and improve the precision to which we estimate noise given fixed experimental resources.
We demonstrate the scalability and performance of our protocol through circuit-level numerical simulations of the entire noise characterisation procedure for the syndrome extraction circuit of a distance-25 surface code with over 1000 qubits.
Our results indicate that detailed noise characterisation methods are scalable to near-term quantum devices.
We release our code in the form of the Julia package QuantumACES.
\end{abstract}

\maketitle

\section{Introduction}\label{sec:introduction}

Noise in quantum devices is the key obstacle to large-scale quantum computation.
Detailed noise characterisation methods are essential for demonstrating the correct performance of prototype quantum computers, and can improve the quantum error correction required for reliable performance.
We therefore seek methods capable of scaling to early fault-tolerant quantum computers~\citep{martinis_qubit_2015}.
Tomographic methods, such as those implemented in~\citep{chuang_prescription_1997, haeffner_scalable_2005}, as well as gate set tomography~\citep{nielsen_gate_2021}, attempt to fully characterise noise processes and scale poorly as a result.
This has motivated a large literature on more efficient noise characterisation methods, including compressed sensing tomography~\citep{gross_quantum_2010} and shadow tomography~\citep{huang_predicting_2020}.
Randomised benchmarking and its variants~\citep{emerson_scalable_2005, magesan_characterizing_2012, magesan_efficient_2012, gambetta_characterization_2012, wallman_randomized_2018, helsen_new_2019, flammia_efficient_2020} are much more efficient but do not characterise noise in detail.
There are a number of related methods that fit a reduced set of noise parameters within a restricted class of noise models, such as Pauli channel estimation techniques~\citep{flammia_efficient_2020, harper_efficient_2020, harper_fast_2021, flammia_pauli_2021,flammia_averaged_2022}, cycle benchmarking~\citep{erhard_characterizing_2019}, and cycle error reconstruction~\citep{carignan-dugas_error_2023}.

Scalable noise characterisation methods can improve the performance of error correction by identifying which physical configurations of errors are more and less likely~\citep{sundaresan_demonstrating_2023}.
Noise characterisation methods can also identify biased noise~\citep{gicev_quantum_2024}, to which codes~\citep{tuckett_ultrahigh_2018, tuckett_tailoring_2019, bonillaataides_xzzx_2021}, decoders~\citep{tuckett_faulttolerant_2020, iolius_performance_2022, higgott_improved_2023, tiurev_correcting_2023}, and fault-tolerant operations~\citep{aliferis_faulttolerant_2008}, can be tailored in order to improve their performance.

Recently, rapid experimental progress has begun to demonstrate the utility of fault tolerance and quantum error correction~\citep{egan_faulttolerant_2021, krinner_realizing_2022, zhao_realization_2022, acharya_suppressing_2023, sivak_realtime_2023, wang_faulttolerant_2024, bluvstein_logical_2024, dasilva_demonstration_2024, acharya_quantum_2024, caune_demonstrating_2024, reichardt_logical_2024, bausch_learning_2024}.
Consequently, there is an increasing need for noise characterisation methods suitable for near-term quantum devices with hundreds of qubits.

In this paper, we develop a scalable noise characterisation protocol suited to characterising the syndrome extraction circuits of quantum error correcting codes by building upon averaged circuit eigenvalue sampling (ACES)~\citep{flammia_averaged_2022}.
ACES is a general framework for scalable noise characterisation experiments that simultaneously estimates the Pauli channels associated with the operation of all gates in a Clifford circuit, with a recent implementation effort~\citep{pelaez_average_2024}.
We demonstrate our protocol on simulated data at scales of over \(1000\) qubits in experimentally-relevant noise parameter regimes.

Our protocol leverages the structure of fault-tolerant logical circuits.
Fault-tolerant quantum computation~\citep{shor_faulttolerant_1996,aliferis_quantum_2006, gottesman_introduction_2010, aliferis_introduction_2013} replaces physical state preparations, gates, and measurements with redundantly encoded logical equivalents that entail regularly measuring the parity checks of a quantum error correcting code.
In a typical fault-tolerant architecture, the syndrome extraction circuits that measure parity checks represent the bulk of the physical qubits and gate operations, rendering them the key target for noise characterisation experiments.
We leverage the fact that the simple structures of the syndrome extraction circuits of topological quantum codes remain similar across code sizes.
As syndrome extraction circuits are representative of fault-tolerant gadgets, we believe our methods are capable of characterising other fault-tolerant gadgets such as magic state distillation~\citep{litinski_magic_2019}.

\begin{figure*}[t!]
    \centering
    \begin{tikzpicture}
    \node (image) at (0,0) {
        \includegraphics[width=0.98\textwidth]{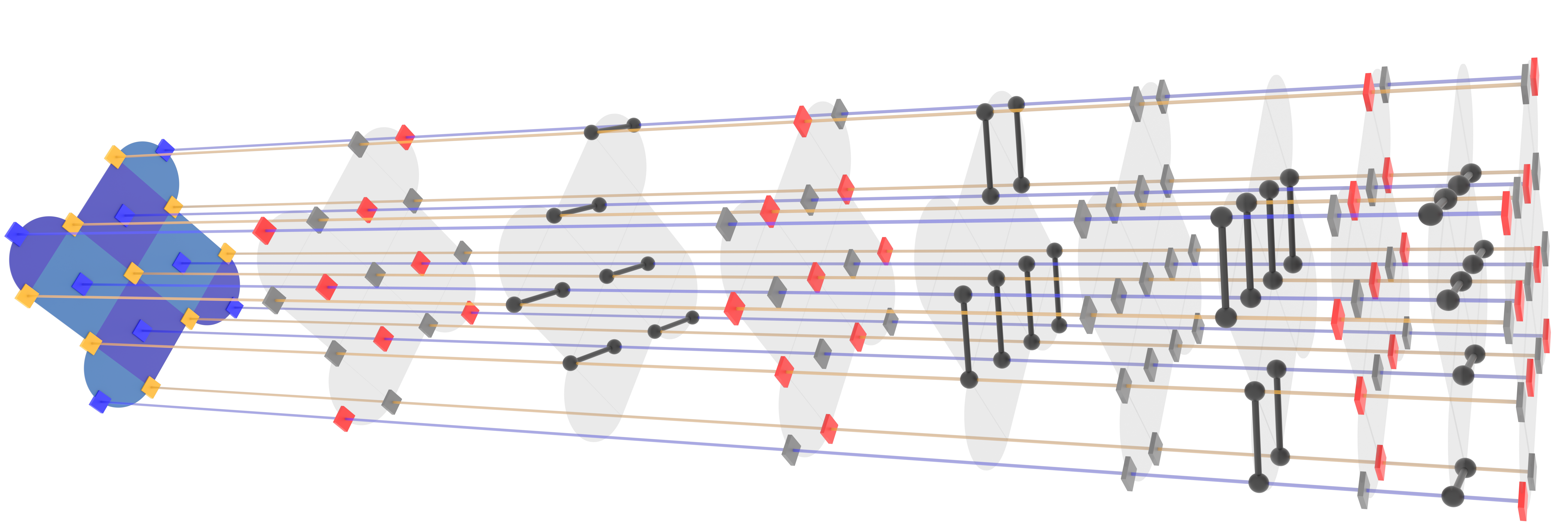}
    };
    \node (A) at (-7.8, 2.8) {Time    };
    \node (B) at (-6.2, 2.8) {};
    \draw [-{Latex[length=2.5mm]}] (A) edge (B);
    \node[draw,circle,inner sep=2pt,font={\footnotesize}] at (-4.50, 2.8) {1};
    \node[draw,circle,inner sep=2pt,font={\footnotesize}] at (-1.91, 2.8) {2};
    \node[draw,circle,inner sep=2pt,font={\footnotesize}] at (0.39,  2.8) {3};
    \node[draw,circle,inner sep=2pt,font={\footnotesize}] at (2.42,  2.8) {4};
    \node[draw,circle,inner sep=2pt,font={\footnotesize}] at (4.08,  2.8) {5};
    \node[draw,circle,inner sep=2pt,font={\footnotesize}] at (5.51,  2.8) {6};
    \node[draw,circle,inner sep=2pt,font={\footnotesize}] at (6.63,  2.8) {7};
    \node[draw,circle,inner sep=2pt,font={\footnotesize}] at (7.60,  2.8) {8};
    \node[draw,circle,inner sep=2pt,font={\footnotesize}] at (8.37,  2.8) {9};
    \end{tikzpicture}
    \caption{Diagram of the syndrome extraction circuit for a distance-\(3\) surface code with \(17\) qubits.
    The circuit implements the XZZX variant of the surface code~\citep{wen_quantum_2003, bonillaataides_xzzx_2021}, and its structure follows a recent experiment~\citep{acharya_suppressing_2023}.
    The code consists of \(9\) data qubits (yellow) and \(8\) measure qubits (blue), and each measure qubit is associated with a coloured plaquette representing the \(Z\) (light blue) or \(X\) (dark blue) stabiliser it measures.
    The circuit is divided into \(9\) numbered gate layers (light grey) implemented sequentially in time, which feature controlled-\(Z\) gates (black), Hadamard gates (red), and dynamical decoupling Pauli \(X\) gates (grey).
    Layers \(1\) and \(9\) are identical, as are layers \(3\) and \(7\).
    We learn a circuit-level noise model, namely single- and two-qubit Pauli channels, depending on the gate, operated in the context of its layer, as well as single-qubit Pauli channels for idle qubits in each layer.}
    \label{fig:rotated-code-circuit}
\end{figure*}
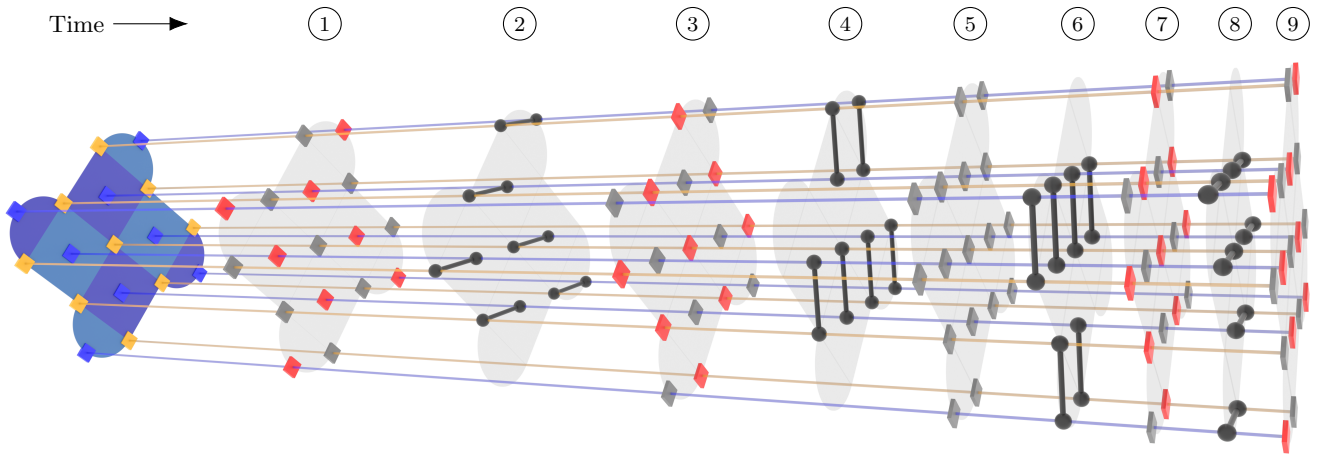

In \cref{sec:characterisation-scale}, we offer a high-level overview of our modified formulation of ACES, which naturally describes scalable and performant noise characterisation of the syndrome extraction circuits of topological quantum error correcting codes such as the surface code~\citep{bravyi_quantum_1998, dennis_topological_2002, kitaev_faulttolerant_2003, fowler_surface_2012}.
A detailed overview follows in \cref{sec:aces-overview}, where we rigorously analyse the performance of ACES experiments to derive a figure of merit for their experimental design as well as an improved noise estimation procedure.
Then \cref{sec:designing-aces} describes the experimental design techniques enabling our protocol, including routines that leverage our figure of merit to optimise experimental designs for the average error rates we expect to encounter in a device.

We present numerical results in \cref{sec:numerical-results} for the surface code syndrome extraction circuit, which demonstrate that the performance of optimised experimental designs is asymptotically independent of the code distance and robust to differing error rates or a different noise model.
We close by validating our methods in circuit-level numerical simulations of the noise characterisation of the syndrome extraction circuit of a distance-\(25\) surface code with \(1249\) qubits, performed on a laptop, and release our code in the form of the Julia package QuantumACES~\citep{hockings_quantumacesjl_2025}.
Finally, we provide concluding remarks in \cref{sec:conclusions}.
Our results demonstrate that ACES is capable of scalable and performant characterisation of circuit-level Pauli noise, enabling greater understanding of quantum devices, and paving the way for noise-aware decoders and tailored quantum error correcting codes.

\section{Noise characterisation at scale}\label{sec:characterisation-scale}

We aim to formulate a protocol, built upon ACES, that naturally describes noise characterisation experiments for the syndrome extraction circuits of topological quantum codes.
We focus on surface codes, which encode a logical qubit into a \(d_Z\times d_X\) rectangular array of physical qubits called \emph{data qubits}.
The stabilisers are mutually commuting \(Z\) and \(X\) parities of clusters of data qubits.
They are measured by a syndrome extraction circuit, which entangles each of the \(d_Zd_X-1\) \emph{measure qubits} with adjacent data qubits.
Recent experimental progress has demonstrated improved average performance of a surface code with increasing distance~\citep{acharya_suppressing_2023, acharya_quantum_2024}.
We will focus on the syndrome extraction circuit used in this experiment, which is depicted in \cref{fig:rotated-code-circuit} for a square (\(d=d_Z=d_X\)) distance-\(3\) surface code.

The syndrome extraction circuit is organised as a series of layers implemented sequentially in time.
Each layer consists of a set of \emph{gates}, relatively primitive operations on the quantum device that act on a bounded \(\mathcal{O}(1)\) number of qubits, independent of the total number of qubits \(n\) in the code.
The \emph{layers} are chosen such that gates in a layer are simultaneously physically implemented by the quantum device and act on disjoint sets of qubits, ensuring that gates in a layer trivially commute with each other.
What we call layers are referred to elsewhere as cycles~\citep{wallman_noise_2016, erhard_characterizing_2019, carignan-dugas_error_2023}, or gates~\citep{nielsen_gate_2021}.
For example, the surface code syndrome extraction circuit depicted in \cref{fig:rotated-code-circuit} is divided into \(9\) layers consisting of controlled-\(Z\) gate layers interspersed between layers of Hadamard and dynamical decoupling \(X\) gates.

Quantum noise can be roughly divided into coherent errors, which interfere with themselves and can accumulate, and incoherent noise, which cannot.
Coherent errors can be mitigated by improved device calibration or methods such as dynamical decoupling~\citep{viola_dynamical_1999}.
They can also be tailored into incoherent Pauli noise by quantum error correction~\citep{beale_coherence_2018, huang_performance_2019, iverson_coherence_2020}, as well as Pauli frame randomisation~\citep{knill_quantum_2005, ware_experimental_2021}, which can for example be realised by the randomised compiling protocol~\citep{wallman_noise_2016, hashim_randomized_2021}.
Indeed, these randomisation techniques can improve the performance of quantum error correction in the presence of coherent errors~\citep{jain_improved_2023}.

We will learn a circuit-level incoherent Pauli noise model by characterising the noise associated with each gate in the circuit when implemented in the context of the layer in which it appears.
This captures averaged spatial correlations between each gate and the other gates in its layer.
We assume that the noise associated with each layer is independent of the prior and subsequent layers, as well as the time at which the layer was implemented.
This type of noise is often referred to in benchmarking as Markovian and time-stationary~\citep{wallman_randomized_2018, flammia_efficient_2020}.
Dynamical decoupling pulses are believed to improve the reliability of this assumption, and we observe that the dynamical decoupling layer \(5\) in \cref{fig:rotated-code-circuit} ensures that all two-qubit gate layers are separated by single-qubit gate layers.
Also, the noise associated with the \(X\) gates in layer \(5\) is learned separately from the noise associated with their counterparts acting on the same qubits in layers \(1\), \(3\), \(7\), and \(9\).

At its core, ACES estimates the noise associated with a set of circuits constructed by rearranging the layers of the original circuit in order to infer the noise associated with each of the gates in the original circuit.
These rearrangements are more precisely permutations with repetition, or \emph{tuples} \(T\), which will allow us to parameterise the design of a noise characterisation experiment by the \emph{tuple set} \(\mathcal{T}\).
Tuples can be of arbitrary length, and their elements are drawn from the \emph{unique layer indices} \(\mathcal{I}\).
For example, in \cref{fig:rotated-code-circuit} layers \(1\) and \(9\) and layers \(3\) and \(7\) are identical, so \(\mathcal{I}=\{1,2,3,4,5,6,8\}\), and the tuple for the original circuit is \(T=(1,2,3,4,5,6,3,8,1)\).

We focus on tuples because the structure of the surface code syndrome extraction circuit is similar regardless of the code size.
This is entailed by two properties, namely that the surface code is a quantum low-density parity-check (LDPC) code, and the stabilisers are geometrically local.
Specifically, the stabilisers of quantum LDPC codes include a constant \(\mathcal{O}(1)\) number of qubits, and their qubits are included in a constant \(\mathcal{O}(1)\) number of stabilisers~\citep{breuckmann_quantum_2021}.
Topological quantum codes have these properties, which allows us to divide their syndrome extraction circuits into a number of layers \(l\), the circuit depth, that is independent of the number of qubits in the code \(n\).

Consequently, we will see that the performance of a tuple set is empirically precisely predictable as a function of the surface code distance \(d\).
This will allow us to optimise tuple sets according to their performance at small code distances, and then use the optimised experimental design to characterise the noise associated with large surface codes for which direct optimisation is infeasible.
Both the noise estimation precision and the number of experiments required on the quantum device are asymptotically independent of the surface code distance \(d\).
This achieves our goal of performance at scale.

ACES is a flexible framework capable of describing noise characterisation experiments that learn more detailed spatial and temporal correlations than we consider here.
However, learning such correlations trades off against the performance and scalability of the protocol.
While our numerical results focus on a syndrome extraction circuit for the \emph{rotated} surface code, which we simply call the surface code, \cref{apdx:unrotated-surface-codes} shows that our methods perform remarkably similarly for the \emph{unrotated} surface code.
We believe that our methods are capable of characterising the noise associated with the syndrome extraction circuits of topological quantum codes, as well as other fault-tolerant gadgets such as magic state distillation~\citep{litinski_magic_2019}.
Practical architectures for quantum LDPC codes will need to address the issue of measuring non-local stabilisers, which may also allow our methods to extend to these codes.

\section{ACES overview}\label{sec:aces-overview}

We now overview ACES, recasting the original presentation of~\citep{flammia_averaged_2022} to instead view circuits as sequences of layers.
This allows us to parameterise noise characterisation experiments by a set of tuples that rearrange these circuit layers, a natural framing for the syndrome extraction circuits of topological quantum codes.

Then we analyse the eigenvalue estimation procedure of~\citep{flammia_direct_2011} in the context of simultaneous estimation of multiple eigenvalues.
This allows us to improve upon the core least squares noise estimation procedure given in~\citep{flammia_averaged_2022}, and derive a figure of merit for the expected performance of ACES noise characterisation experiments which we later use to optimise their experimental design.

\subsection{Mathematical preliminaries}\label{sec:mathematical-preliminaries}

We begin with a few mathematical preliminaries.
The \(n\)-qubit \emph{Pauli group} \(\mathbf{P}^n\) contains all \(n\)-fold tensor products of Pauli operators \(P_{\bm{a}}\) indexed by length \(2n\) bit strings \(\bm{a}=(\bm{a}^{(x)},\bm{a}^{(z)})=(a_{1}^{(x)},\dots,a_{n}^{(x)},a_{1}^{(z)},\dots,a_{n}^{(z)})\) such that
\begin{equation}\label{eq:pauli}
    P_{\bm{a}}=\bigotimes_{j=1}^{n}{i^{a_{j}^{(x)}a_{j}^{(z)}}X^{a_{j}^{(x)}}Z^{a_{j}^{(z)}}}.
\end{equation}
Indeed, the group contains all \(\zeta P_{\bm{a}}\), where \(\zeta\in\{\pm 1,\pm i\}\) is an overall phase factor, though we generally write \(P_{\bm{a}}\in\mathbf{P}^n\) to refer to the unit phase element of the group.
The group operation is matrix multiplication, and corresponds, up to a phase factor, to bit string addition
\begin{equation}\label{eq:pauli-operation}
    P_{\bm{a}}P_{\bm{a}^\prime}=\zeta(\bm{a},\bm{a}^\prime)P_{\bm{a}+\bm{a}^\prime}.
\end{equation}
It is therefore convenient to refer to Paulis \(P_{\bm{a}}\) by their bit string \(\bm{a}\), and we will often write \(\bm{a}\in\mathbf{P}^n\).
The group commutation relations are given by the symplectic bilinear form \(\omega(\bm{a},\bm{a}^\prime)=\bm{a}^{(x)}\cdot\bm{a}^{\prime(z)}+\bm{a}^{(z)}\cdot\bm{a}^{\prime(x)}\) as
\begin{equation}\label{eq:pauli-commute}
    P_{\bm{a}}P_{\bm{a}^\prime}={(-1)}^{\omega(\bm{a},\bm{a}^\prime)} P_{\bm{a}^\prime}P_{\bm{a}},
\end{equation}
which is also symmetric as we work over \(\mathbb{Z}_2\).
Finally, define the \emph{support} of a Pauli \(\bm{a}\) to be the set of qubits on which that Pauli is not the identity, that is,
\begin{equation}\label{eq:pauli-support}
    \supp{(\bm{a})}=\{j\in[n]\colon\{a_{j}^{(x)},a_{j}^{(z)}\}\ne\{0,0\}\},
\end{equation}
where we adopt the convention \([n]=\{1,\dots,n\}\).
A Pauli supported on \(b\) qubits is also called a weight-\(b\) Pauli.

A \emph{Pauli channel} (for example, see~\citep{fawzi_lower_2023}) is a quantum channel that can be represented in the form
\begin{equation}\label{eq:pauli-channel}
    \mathcal{E}(\rho)=\sum_{\bm{a}\in\mathbf{P}^n}{p_{\bm{a}}P_{\bm{a}}\rho P_{\bm{a}}},
\end{equation}
where the \emph{Pauli error probabilities} \(p_{\bm{a}}\) form a probability distribution over Pauli operators.
We seek to characterise noise represented in the form of Pauli channels by estimating the Pauli error probabilities of those channels.
To this end, consider the \emph{Pauli channel eigenvectors}, which are simply the Pauli operators
\begin{equation}\label{eq:pauli-eigenvectors}
    \mathcal{E}(P_{\bm{a}^\prime})=\lambda_{\bm{a}^\prime}P_{\bm{a}^\prime}.
\end{equation}
From \cref{eq:pauli-commute,eq:pauli-channel,eq:pauli-eigenvectors}, we see that the \emph{Pauli channel eigenvalues} \(\lambda_{\bm{a}^\prime}\) are related to the Pauli error probabilities \(p_{\bm{a}}\) as
\begin{equation}\label{eq:pauli-eigenvalues-from-probabilities}
    \lambda_{\bm{a}^\prime}=\sum_{\bm{a}\in\mathbf{P}^n}{{(-1)}^{\omega(\bm{a},\bm{a}^\prime)}p_{\bm{a}}}.
\end{equation}
In fact, this relation is a \emph{Walsh-Hadamard transform}, ordered according to the symplectic form \(\omega\) rather than the usual inner product~\citep{harper_fast_2021}.
As the transform is its own inverse, up to a constant, the error probabilities can similarly be determined from the eigenvalues as
\begin{equation}\label{eq:pauli-probabilities-from-eigenvalues}
    p_{\bm{a}}=\frac{1}{4^n}\sum_{\bm{a}^\prime\in\mathbf{P}^n}{{(-1)}^{\omega(\bm{a},\bm{a}^\prime)}\lambda_{\bm{a}^\prime}}.
\end{equation}
We will focus on estimating the Pauli channel eigenvalues, as they recover the Pauli error probabilities and are easier to estimate.

Define the \(n\)-qubit \emph{Clifford group} \(\mathbf{C}^n\) to be the group generated by the controlled-\(X\), Hadamard, and phase gates, written \(C_i(X_j)\), \(H_j\), and \(S_j\), respectively, for control qubits \(i\) and target qubits \(j\ne i\)~\citep{aaronson_improved_2004}.
Elements of the Clifford group \(G\in\mathbf{C}^n\) act by conjugation on Pauli group elements \(P_{\bm{a}}\) as \(GP_{\bm{a}}G^\dagger={(\pm)}_{G,\bm{a}}P_{G(\bm{a})}\), where the resulting Pauli \(P_{G(\bm{a})}\) and sign \({(\pm)}_{G,\bm{a}}\) can be computed efficiently in the number of qubits \(n\)~\citep{gottesman_stabilizer_1997, aaronson_improved_2004}.
We find it convenient to refer to \(\mathcal{G}\) rather than \(G\), where \(\mathcal{G}(P_{\bm{a}})=GP_{\bm{a}}G^\dagger\), and do so henceforth.
In the context of surface code syndrome extraction circuits, we also encounter the generating gates of the Pauli group \(X_j\), \(Y_j\), and \(Z_j\), as well as the controlled-
\(Z\) gate \(C_i(Z_j)=H_jC_i(X_j)H_j\).
Analogously to the support of a Pauli, define the support of a Clifford group element \(\supp{(\mathcal{G})}\) to be the set of qubits on which that element acts non-trivially.

Specifically, we will be interested in characterising Pauli noise channels \(\mathcal{E}_{\mathcal{G}}\) associated with Clifford gates \(\mathcal{G}\), whose noisy implementation we write as \(\tilde{\mathcal{G}}=\mathcal{G}\mathcal{E}_{\mathcal{G}}\).
The noisy implementation therefore acts as
\begin{equation}\label{eq:clifford-noisy}
    \tilde{\mathcal{G}}(P_{\bm{a}})=\lambda_{\mathcal{G},\bm{a}}\mathcal{G}(P_{\bm{a}})={(\pm)}_{\mathcal{G},\bm{a}}\lambda_{\mathcal{G},\bm{a}}P_{\mathcal{G}(\bm{a})},
\end{equation}
where \(\lambda_{\mathcal{G},\bm{a}}\) is called the \emph{generalised eigenvalue} of \(\mathcal{G}\) acting on \(P_{\bm{a}}\), associated with its Pauli noise channel \(\mathcal{E}_{\mathcal{G}}\).
For convenience, we will refer to generalised eigenvalues as eigenvalues.

\subsection{ACES fundamentals}\label{sec:aces-fundamentals}

We can now present the core idea behind ACES, omitting for clarity the Pauli frame randomisation that renders it `averaged'.
Pauli frame randomisation tailors arbitrary noise channels on average into Pauli channels~\citep{knill_quantum_2005, ware_experimental_2021}, and this can be realised in arbitrary quantum circuits by the randomised compiling protocol~\citep{wallman_noise_2016, hashim_randomized_2021}.
However, as we only consider Clifford circuits, generic Pauli frame randomisation suffices here, and consequently we assume that all noise channels are Pauli channels.
Adding the random Pauli gates required for Pauli frame randomisation to our approach is straightforward, and we refer the reader to the original presentation of ACES for details~\citep{flammia_averaged_2022}.

In our presentation of ACES, we focus on characterising the Pauli noise associated with implementing a \emph{circuit} \(\mathcal{C}\) of Clifford gates acting on \(n\) qubits.
The circuit is divided into a series of \(l\) \emph{layers} \(\mathcal{C}_i\) each composed of some number \(k_i\) of Clifford gates \(\mathcal{G}_{ij}\).
As we assume that the noise is time-independent, the noisy implementation of each layer can be written as \(\tilde{\mathcal{C}}_i=\mathcal{C}_i\mathcal{E}_i\) for some Pauli noise channel \(\mathcal{E}_{i}\).

As it is generally intractable to learn all \(4^n\) eigenvalues of such an \(n\)-qubit Pauli channel, we learn a circuit-level noise model.
This noise model does not explicitly represent noise that is spatially correlated between gates in a layer and only learns up to weight-two Pauli errors because we only consider single- and two-qubit gates.
Specifically, we learn the Pauli channels \(\mathcal{E}_{ij}\) associated with the gates \(\mathcal{G}_{ij}\) operated in the context of each unique layer \(\mathcal{C}_i\) for \(i\in\mathcal{I}\), as well as single-qubit Pauli channels for idle qubits in each layer.
We do this by estimating the \emph{gate eigenvalues} \(\lambda_{ij,\bm{a}}\), which yield the Pauli error probabilities of each channel by \cref{eq:pauli-probabilities-from-eigenvalues}.
In \cref{apdx:estimate-marginal}, we show that the Pauli error probabilities estimated by ACES for the gate channel \(\mathcal{E}_{ij}\) are the marginal of the Pauli error probabilities of the layer channel \(\mathcal{E}_{i}\) onto the support of the gate \(\mathcal{G}_{ij}\), demonstrating that ACES captures averaged spatial correlations between gates appearing in the same layer.

To specify the entire collection of gate eigenvalues, let us first write the set of Pauli operators supported on some gate \(\mathcal{G}\) as
\begin{equation}\label{eq:pauli-supported-set}
    \Pauli{(\mathcal{G})}=\{\bm{a}\in\mathbf{P}^n\colon\supp{(\bm{a})}\subseteq\supp{(\mathcal{G})}\},
\end{equation}
a subgroup of \(\mathbf{P}^n\).
For a single-qubit gate, this consists of the Paulis \(I\), \(X\), \(Y\), and \(Z\), and in general for a gate supported on \(b\) qubits, there are \(4^b\) Paulis in this set.
Then the set of gate eigenvalue indices \(N\) is
\begin{equation}\label{eq:gate-eigenvalue-indices}
    N=\left\{(ij,\bm{a})\colon i\in \mathcal{I},j\in[k_i],\bm{a}\in\Pauli{(\mathcal{G}_{ij})}\right\},
\end{equation}
where \(i\in\mathcal{I}\) indexes the unique layers in the circuit and \(j\in [k_i]\) indexes the gates in the layer \(\mathcal{C}_i\).
Then writing the multi-index \(\nu=(ij,\bm{a})\), we seek to estimate \(\lambda_\nu\) for all \(\nu\in N\).
For trace-preserving channels where leakage is absent, identity eigenvalues are always 1 and do not need to be estimated, so for \(b\)-qubit gates we need only estimate \(4^b-1\) gate eigenvalues.

Next, let us formally introduce the concept of a tuple.
Write the circuit as
\begin{equation}\label{eq:circuit}
    \mathcal{C}=\mathcal{C}_l\dots\mathcal{C}_2\mathcal{C}_1,\quad\mathcal{C}_i=\prod_{j\in[k_i]}{\mathcal{G}_{ij}},
\end{equation}
where we time-order the product for the circuit \(\mathcal{C}\), with \(\mathcal{C}_1\) indicating the first layer and earliest time.
The gates \(\mathcal{G}_{ij}\) in the layer \(\mathcal{C}_i\) commute, so the order of the product does not matter.
Then a \emph{tuple} \(T\) of some arbitrary length \(L\) is a sequence of \(L\) numbers in \(\mathcal{I}\) ordering the layers of the circuit \(\mathcal{C}\) that acts to rearrange it as
\begin{equation}\label{eq:circuit-arrangement}
    \mathcal{C}_{T}=\prod_{u\in[L]}{\mathcal{C}_{T_u}}.
\end{equation}
\Cref{fig:tuple-indexing} gives a concrete example of a three-layer circuit rearranged by the length 4 tuple \(T=(2,1,3,2)\).

\renewcommand{\push}[1]{\color{gray}{#1}}
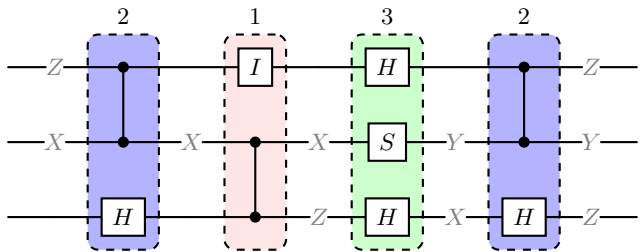
\begin{figure}[t]
    \centering
    \begin{quantikz}[column sep = 0.475cm]
        &\push{Z}& \ctrl{1}\gategroup[3,steps=1,style={dashed,rounded corners,fill=blue!30,inner sep=2pt},background]{2} && \gate{I}\gategroup[3,steps=1,style={dashed,rounded
        corners,fill=red!10,inner sep=2pt},background]{1} && \gate{H}\gategroup[3,steps=1,style={dashed,rounded
        corners,fill=green!20,inner sep=2pt},background]{3} && \ctrl{1}\gategroup[3,steps=1,style={dashed,rounded
        corners,fill=blue!30,inner sep=2pt},background]{2} &\push{Z}& \\
        &\push{X}& \control{} &\push{X}& \ctrl{1} &\push{X}& \gate{S} &\push{Y}& \control{} &\push{Y}& \\
        && \gate{H} && \control{} &\push{Z}& \gate{H} &\push{X}& \gate{H} &\push{Z}& 
    \end{quantikz}
    \caption{Example three-qubit circuit arranged by the tuple \(T=(2,1,3,2)\).
    Each of the Clifford gates \(\mathcal{G}_{ij}\) in the circuit is associated with gate eigenvalues \(\lambda_{ij,\bm{a}}\), where \(i\) labels the layer, \(j\) labels gates within a layer sequentially from the top, and \(\bm{a}\) labels Paulis supported on the gate.
    For example, \(\lambda_{32,X}\) refers to the \(X\) eigenvalue of the phase gate in layer \(3\), and \(\lambda_{21,ZX}\) refers to the \(ZX\) eigenvalue of the controlled-\(Z\) gate in layer \(2\).
    The gates in layer \(2\) have the same gate eigenvalues in both occurrences.
    Layer \(1\) is padded with an identity gate.
    The goal of ACES is to estimate these gate eigenvalues.
    Labels show the propagation of the Pauli \(ZXI\) through the circuit.
    The corresponding circuit eigenvalue is \(\Lambda_{T,ZXI}=\lambda_{21,ZX}\lambda_{12,XI}\lambda_{32,X}\lambda_{33,Z}\lambda_{21,IY}\lambda_{22,X}\).
    To estimate the circuit eigenvalue, prepare eigenstates of \(ZXI\), perform the circuit, and measure in the \(ZYZ\) basis.
    Account for state preparation and measurement (SPAM) errors by including eigenvalues in the circuit eigenvalue product that correspond to each of the three measurements.}
    \label{fig:tuple-indexing}
\end{figure}

Now we are ready to describe the circuit eigenvalues from which we estimate the gate eigenvalues.
Following \cref{eq:clifford-noisy}, the \emph{circuit eigenvalue} \(\Lambda_{T,\bm{a}}\) of the noisy implementation \(\tilde{\mathcal{C}}_T\) of the circuit \(\mathcal{C}_T\) describes the overall effect of the noise on the Pauli \(P_{\bm{a}}\) as
\begin{equation}\label{eq:circuit-action}
    \tilde{\mathcal{C}}_T(P_{\bm{a}})=\Lambda_{T,\bm{a}}\mathcal{C}_T(P_{\bm{a}})={(\pm)}_{T,\bm{a}}\Lambda_{T,\bm{a}}P_{T(\bm{a})}.
\end{equation}
For convenience, we write \(P_{T(\bm{a})}\) to refer to \(P_{\mathcal{C}_T(\bm{a})}\), and similarly with the sign.

In fact, \cref{eq:circuit-action} is identical to \cref{eq:clifford-noisy}, except that the circuit \(\mathcal{C}_T\) can be decomposed into layers of gates, each with their own gate eigenvalues.
Tracking the evolution of the Pauli as it is acted upon by the circuit allows us to determine how the circuit eigenvalue is composed as a product of gate eigenvalues, as shown in \cref{fig:tuple-indexing}.
Using \(u\) to index the circuit layer time-step, write \(P_{\bm{a}^{(u)}}\) for the state of the Pauli before the action of the \(u\)th layer, so that \(\mathcal{C}_{T_u}\) maps \(P_{\bm{a}^{(u)}}\) to \(P_{\bm{a}^{(u+1)}}\), up to a sign factor.
Then a simple calculation yields
\begin{equation}\label{eq:circuit-eigenvalue-timeseries}
    \Lambda_{T,\bm{a}}=\prod_{u\in[L]}{\prod_{j\in[k_{T_u}]}{\lambda_{T_{u}j,\bm{a}^{(u)}}}},
\end{equation}
where the gate eigenvalues \(\lambda_{T_{u}j,\bm{a}^{(u)}}\) implicitly depend only on the qubits of \(P_{\bm{a}^{(u)}}\) which lie in the support of the gates \(\mathcal{G}_{T_{u}j}\) of the \(u\)th layer \(\mathcal{C}_{T_u}\).

The key insight of ACES is that the relation \cref{eq:circuit-eigenvalue-timeseries} can be inverted, allowing us to estimate all of the gate eigenvalues using estimates of a sufficiently large and varied set of circuit eigenvalues.
For a \emph{tuple set} \(\mathcal{T}=\{T\}\) to specify this set of circuit eigenvalues and hence the experimental design, we must supply a rule for choosing the circuit eigenvalues estimated for a tuple \(T\).
We will assume that we can prepare qubits in each of the Pauli \(X\), \(Y\), and \(Z\) bases, which usually requires a small number of single-qubit Clifford gates whose associated noise we do not explicitly characterise.
Allowing only \(Z\) basis preparations would constrain the circuit and gate eigenvalues we are able to estimate~\citep{huang_foundations_2022}.
Specifically, we will estimate circuit eigenvalues \(\Lambda_{T,\bm{a}}\) corresponding to all of the Paulis \(\bm{a}\in\Pauli{(\mathcal{G})}\) supported on some gate \(\mathcal{G}\in\mathcal{C}_T\) in the rearranged circuit.

This rule ensures that there exists at least one tuple set whose circuit eigenvalues are sufficient for estimating all of the gate eigenvalues.
Namely, consider the tuple set \(\mathcal{T}_{\mathcal{I}}^{\prime}=\{(i)\}_{i\in\mathcal{I}}\) whose tuples are each of the unique layer indices \(\mathcal{I}\) of the circuit repeated just once.
This tuple set corresponds to direct estimation of the gate eigenvalues, as the circuit eigenvalues are precisely the gate eigenvalues and the design matrix is the identity up to the ordering of the circuit and gate eigenvalues.

Accordingly, define the \emph{Pauli preparation set} \(\mathcal{Q}_T\) as
\begin{equation}\label{eq:pauli-preparation-set}
    \mathcal{Q}_T=\bigcup_{\mathcal{G}\in\mathcal{C}_T}{\Pauli{(\mathcal{G})}}.
\end{equation}
Then the set of circuit eigenvalue indices \(M\) is
\begin{equation}\label{eq:circuit-eigenvalue-indices}
    M=\{(T,\bm{a})\colon T\in\mathcal{T},\bm{a}\in\mathcal{Q}_T\}.
\end{equation}
Hence, writing the multi-index \(\mu=(T,\bm{a})\), we will estimate the circuit eigenvalues \(\Lambda_\mu\) for all \(\mu\in M\).

Now rewrite the product over circuit layer time-steps \cref{eq:circuit-eigenvalue-timeseries} as a product over the gate eigenvalue indices
\begin{equation}\label{eq:circuit-eigenvalue-design}
    \Lambda_{\mu}=\prod_{\nu\in N}{\lambda_{\nu}^{A_{\mu\nu}}},
\end{equation}
where \(A_{\mu\nu}\) is the non-negative integer power of \(\lambda_\nu\) as it appears in the expression for \(\Lambda_\mu\).
As we are interested in circuit and gate eigenvalues that are not too small, lying in the interval \((\delta,1]\) for some arbitrary \(\delta>0\), we can transform the product of \cref{eq:circuit-eigenvalue-design} into a sum by taking the (negative) logarithm.
This yields circuit and gate (negative) log-eigenvalues \(b_\mu=-\log{\Lambda_{\mu}}\) and \(x_\nu=-\log{\lambda_{\nu}}\) which correspondingly lie in \([0, \log{(1/\delta)})\).
Constructing the \emph{design matrix} \(A\) elementwise from the \(A_{\mu\nu}\) for each \(\mu\in M\) and \(\nu\in N\) obtains
\begin{equation}\label{eq:design-matrix}
    b_{\mu}=\sum_{\nu\in N}{A_{\mu\nu}x_{\nu}}.
\end{equation}
This is a standard linear regression problem, so we can estimate the gate log-eigenvalues from the circuit log-eigenvalues with a linear least squares method as long as the design matrix \(A\), which is usually sparse, has full rank \(N\).
Note that we have overloaded \(N\) to refer both to the gate eigenvalue index set as well as its size, and do the same for the circuit eigenvalue index set \(M\).

We also include gate eigenvalue parameters to account for state preparation and measurement (SPAM) errors.
Owing to a gauge freedom, preparation and measurement errors cannot be estimated separately~\citep{nielsen_gate_2021}, a phenomenon that has also been analysed in the context of Pauli noise learning~\citep{chen_efficient_2024}.
We will not treat this issue in detail, instead making the simple assumption that SPAM noise is fully associated with measurements, effectively fixing a gauge.
Therefore, we model Pauli frame randomised noisy measurements with a measurement error probability \(p_{\nu}\) corresponding to an eigenvalue \(\lambda_{\nu}=1-2p_{\nu}\).
Here, we learn measurements in the Pauli \(X\), \(Y\), and \(Z\) bases separately, yielding \(3n\) SPAM parameters.
We could also assume that the measurement errors are the same across Pauli bases, yielding \(n\) SPAM parameters, but choose not to do so here.

With the addition of SPAM errors, the aforementioned tuple set \(\mathcal{T}_{\mathcal{I}}^{\prime}=\{(i)\}_{i\in\mathcal{I}}\) no longer produces a full-rank design matrix.
We therefore augment it with the empty tuple \(()\) to measure SPAM noise, obtaining the \emph{basic tuple set} \(\mathcal{T}_{\mathcal{I}}=\{(i)\}_{i\in\mathcal{I}}\cup\{()\}\).
This tuple set constructs a full-rank design matrix for any circuit \(\mathcal{C}\).
We describe an algorithm for generating better-performing tuple sets in \cref{sec:tuple-optimisation}.

As we discussed in \cref{sec:characterisation-scale}, the number of layers \(l\) in the syndrome extraction circuits of topological quantum codes is independent of the number of qubits in the code \(n\).
Since gates act on a bounded number of qubits, each gate has a bounded number of gate eigenvalues also independent of \(n\), so we need only estimate \(\mathcal{O}(n)\) gate eigenvalues.
The basic tuple set illustrates that this only requires us to estimate \(\mathcal{O}(n)\) circuit eigenvalues.

\subsection{Circuit eigenvalue sampling}\label{sec:circuit-eigenvalue-sampling}

In this section, we analyse the performance of the eigenvalue sampling procedure of~\citep{flammia_direct_2011}, with a focus on simultaneous estimation of circuit eigenvalues.
We aim to estimate the noise associated with the syndrome extraction circuit \(\mathcal{C}\) in a number of experiments that is asymptotically independent of the number of qubits \(n\) on which the circuit acts.
As we need to estimate \(\mathcal{O}(n)\) circuit eigenvalues, each individual experiment must also estimate \(\mathcal{O}(n)\) circuit eigenvalues.
As Pauli preparations and measurements supported on disjoint sets of qubits can trivially be performed simultaneously, this procedure will allow us to estimate the requisite circuit eigenvalues in \(\mathcal{O}(1)\) experiments.

First, we analyse the estimation of a single circuit eigenvalue.
Examining \cref{eq:circuit-action} and noting that the Pauli operators form an orthogonal basis under the trace inner product, we obtain
\begin{equation}\label{eq:trace-eigenvalue}
    \Lambda_{T,\bm{a}}={(\pm)}_{T,\bm{a}}\frac{1}{2^n}\tr{\left(P_{T(\bm{a})}\tilde{\mathcal{C}}_T(P_{\bm{a}})\right)}.
\end{equation}
This immediately suggests a sampling procedure for estimating the circuit eigenvalue \(\Lambda_{T,\bm{a}}\), where we prepare eigenstates of \(P_{\bm{a}}\), apply the noisy circuit \(\tilde{\mathcal{C}}_T\), and then measure in the Pauli basis defined by \(P_{T(\bm{a})}\), correcting the outcome according to the sign \({(\pm)}_{T,\bm{a}}\).

Specifically, consider the eigenbasis \(|\psi_{\bm{s}}^{\bm{a}}\rangle\) of \(P_{\bm{a}}\) consisting of all \(2^n\) sign configurations of single-qubit Pauli eigenstate tensor products, indexed by the length \(n\) bit string \(\bm{s}\).
The state \(|\psi_{\bm{s}}^{\bm{a}}\rangle\) is an eigenstate of \(P_{\bm{a}}\) with eigenvalue \(s\), the parity of the vector \(\bm{s}\).

Suppose that we prepare such an eigenstate, with a sign configuration \(\bm{s}\) chosen uniformly at random, implement the noisy circuit \(\tilde{\mathcal{C}}_T\), and measure in the basis defined by \(P_{T(\bm{a})}\), correcting the outcome for the signs \({(\pm)}_{T,\bm{a}}\) and \(s\).
The measurement outcome \(m_{T,\bm{a}}\) is a Bernoulli random variable that takes values \(\pm 1\), and \cref{eq:trace-eigenvalue} shows that it has expectation \(\Lambda_{T,\bm{a}}\) and hence variance \(1-\Lambda_{T,\bm{a}}^2\).
The sample average estimator \(\hat{\Lambda}_{T,\bm{a}}\) is therefore an unbiased estimator of the circuit eigenvalue \(\Lambda_{T,\bm{a}}\).

We aim to simultaneously estimate multiple circuit eigenvalues in a single experiment.
In general, we could prepare eigenstates of two commuting Paulis \(\bm{a}\) and \(\bm{a}^\prime\), and then measure in the commuting bases generated by \(T(\bm{a})\) and \(T(\bm{a}^\prime)\).
However, these preparations and measurements would generally require noisy entangling gates.
As we cannot fully characterise SPAM noise, we restrict ourselves to individually preparing and measuring each qubit in the Pauli \(X\), \(Y\), or \(Z\) basis, which requires only single-qubit Clifford gates.

This imposes a stricter requirement on simultaneously estimable circuit eigenvalues.
We say that two Paulis \(\bm{a}\) and \(\bm{a}^\prime\) are \(T\)-\emph{consistent}, written \(\bm{a}\tconsistent\bm{a}^\prime\), if on each qubit of the preparations as well as the measurements, either both Paulis are the same or at least one is the identity, such that we can simultaneously estimate the circuit eigenvalues \(\Lambda_{T,\bm{a}}\) and \(\Lambda_{T,\bm{a}^\prime}\).
More formally, \(\bm{a}\tconsistent\bm{a}^\prime\) if \(\bm{a}_j=\bm{a}^\prime_j\) for all \(j\in\supp{(\bm{a})}\cap\supp{(\bm{a}^\prime)}\), and \(T(\bm{a})_j=T(\bm{a}^\prime)_j\) for all \(j\in\supp{(T(\bm{a}))}\cap\supp{(T(\bm{a}^\prime))}\).

Then an \emph{experiment} is a set of mutually \(T\)-consistent Pauli preparations, a subset \(E_T\subset\mathcal{Q}_T\) such that \(\bm{a}\tconsistent\bm{a}^\prime\) for all \(\bm{a},\bm{a}^\prime\in E_T\).
We describe an algorithm for generating experiments in \cref{sec:packing-sets}.
Since all of the preparations are \(T\)-consistent, there is a unique non-identity Pauli on each qubit of the preparations and the measurements, allowing us to easily derive the corresponding \(n\)-qubit preparation \(P_{\bm{A}}\) and measurement \(P_{\bm{A}^\prime}\).
We write the measurement as \(P_{\bm{A}^\prime}\) because in general \(P_{\bm{A}^\prime}\ne P_{T(\bm{A})}\).

As before, randomly prepare some eigenstate \(|\psi_{\bm{s}}^{\bm{A}}\rangle\) of \(P_{\bm{A}}\), implement the circuit \(\tilde{\mathcal{C}}_T\), and then measure each qubit in the basis defined by \(P_{\bm{A}^\prime}\).
This yields a length \(n\) bit string outcome \(\bm{m}_{T,\bm{A}}\), which we correct overall by the sign \({(\pm)}_{T,\bm{A}}\) and elementwise by \(\bm{s}\).
Then the parities \(m_{T,\bm{a}}\) of the marginals of \(\bm{m}_{T,\bm{A}}\) to \(\supp{(T(\bm{a}))}\) contribute to the sample average estimators \(\hat{\Lambda}_{T,\bm{a}}\) for all \(\bm{a}\in E_T\).

Simultaneous measurement predictably correlates the circuit eigenvalue estimators.
Consider the product of measurement outcomes \(m_{T,\bm{a}}m_{T,\bm{a}^\prime}\) for \(\bm{a},\bm{a}^\prime\in E_T\).
As \(P_{\bm{a}}P_{\bm{a}^\prime}=P_{\bm{a}+\bm{a}^\prime}\), this is the measurement outcome of \(\bm{a}+\bm{a}^\prime\), with circuit eigenvalue \(\Lambda_{T,\bm{a}+\bm{a}^\prime}\).
Then the covariance is
\begin{equation}\label{eq:measurement-covariance}
    \cov{[m_{T,\bm{a}},m_{T,\bm{a}^\prime}]}=\Lambda_{T,\bm{a}+\bm{a}^\prime}-\Lambda_{T,\bm{a}}\Lambda_{T,\bm{a}^\prime},
\end{equation}
which reduces to the variance \(1-\Lambda_{T,\bm{a}}^2\) when \(\bm{a}=\bm{a}^\prime\).
Measurement outcomes can only covary within each experiment of the \emph{experiment set} \(\mathcal{E}_T=\{E_T\}\) for a tuple \(T\), which estimates all circuit eigenvalues \(\Lambda_{T,\bm{a}}\) for \(\bm{a}\in\mathcal{Q}_T\).

When characterising noise on a quantum device, we aim to maximise the precision of our estimation given a fixed amount of time on the device.
The time allotted to noise characterisation will be informed by factors including the timescale of device drift, the cost of device time, and the desired duty cycle balance between device characterisation and operation.
Longer tuples and deeper circuits take longer to perform, so the length of the tuples in the tuple set trades off against the number of measurement shots we are able to collect.

Therefore, let us introduce some formalism to define the \emph{measurement budget} \(S\), a modified version of the \emph{measurement shots} \(S^\prime\) that accounts for the time taken to implement the circuits on the device.
Suppose that the circuit corresponding to each tuple \(T\) takes some time \(\tau_T\) to implement on the device.
We allocate a fraction \(\Gamma_T\) of the shots \(S^\prime\) to each experiment set \(\mathcal{E}_T\), where the non-negative \emph{shot weights} \(\Gamma=\{\Gamma_T\}_{T\in\mathcal{T}}\) sum to unity, and divide shots evenly between the individual experiments in each experiment set.
By default, we choose the shot weights \(\Gamma_T\propto\tau_T^{-1}\) such that the experiments for each tuple are measured for an equal amount of time on the device.
Now define the \emph{time factor} \(\tau_{\mathcal{T},\Gamma}=\sum_{T\in\mathcal{T}}{\Gamma_T\tau_T}\), where for the default shot weights we omit the \(\Gamma\) dependence to write \(\tau_{\mathcal{T}}\).
Then the measurement budget \(S\) is given by 
\begin{equation}\label{eq:measurement-budget}
    S=S^\prime\left(\frac{\tau_{\mathcal{T}_{\mathcal{I}}}}{\tau_{\mathcal{T},\Gamma}}\right),
\end{equation}
and represents the number of measurement shots collected by the ACES \emph{experimental design} \((\mathcal{T},\Gamma)\) in the time the basic tuple set collects \(S^\prime\) measurement shots.
Lastly, let the \emph{experiment measurement budget} \(S_T=S\Gamma_T/|\mathcal{E}_T|\) be the measurement budget for each experiment in the experiment set \(\mathcal{E}_T\) for the tuple \(T\).

Finally, we are ready to specify the elements of the \emph{circuit eigenvalue estimator covariance matrix} \(\Omega\) of the circuit eigenvalue estimator vector \(\hat{\bm{\Lambda}}\) by modifying \cref{eq:measurement-covariance} to account for the number of experiments measuring the relevant circuit eigenvalues and the experiment duration.
Let \(E_{T,\bm{a}}\) be the number of experiments \(E_T\in\mathcal{E}_T\) measuring the circuit eigenvalue \(\Lambda_{T,\bm{a}}\), and similarly let \(E_{T,\bm{a},\bm{a}^\prime}\) be the number of experiments that simultaneously measure \(\Lambda_{T,\bm{a}}\) and \(\Lambda_{T,\bm{a}^\prime}\), noting that \(E_{T,\bm{a},\bm{a}}=E_{T,\bm{a}}\).
Then each eigenvalue is measured with \(S_{T}E_{T,\bm{a}}\) shots, and \(\Lambda_{T,\bm{a}}\) and \(\Lambda_{T,\bm{a}^\prime}\) are simultaneously measured with \(S_{T}E_{T,\bm{a},\bm{a}^\prime}\) shots.
This yields the expression for the covariance matrix entries
\begin{equation}\label{eq:eigenvalue-covariance}
    \cov{\!\left[\hat{\Lambda}_{T,\bm{a}},\hat{\Lambda}_{T,\bm{a}^\prime}\right]}
    =\frac{E_{T,\bm{a},\bm{a}^\prime}}{S_{T}E_{T,\bm{a}}E_{T,\bm{a}^\prime}}\big(\Lambda_{T,\bm{a}+\bm{a}^\prime}-\Lambda_{T,\bm{a}}\Lambda_{T,\bm{a}^\prime}\big).
\end{equation}
Note that the covariance matrix is block diagonal, as the circuit eigenvalue estimators for different tuples are uncorrelated, and sparse, as entries are zero when the supports of \(\bm{a}\) and \(\bm{a}^\prime\) remain disjoint throughout their evolution under \(\mathcal{C}_T\).

\subsection{Estimating gate Pauli errors}\label{sec:estimating-pauli-errors}

We now complete the ACES noise estimation procedure by describing a method for estimating the Pauli error probabilities of each of the gates from estimates of the circuit eigenvalues.
This primarily entails solving the linear regression problem posed in \cref{eq:design-matrix} to estimate the gate log-eigenvalues from the circuit log-eigenvalues.
We use \cref{eq:eigenvalue-covariance} for the circuit eigenvalue estimator covariance matrix \(\Omega\) to replace the ordinary least squares method described in~\citep{flammia_averaged_2022} with weighted least squares.
Weighted least squares offers greater performance than ordinary least squares while remaining practically scalable past a thousand qubits.
We further discuss and compare least squares methods in \cref{apdx:least-squares}.

Begin by casting the linear regression problem of \cref{eq:design-matrix} into the standard form
\begin{equation}\label{eq:least-squares}
    \bm{b}=A\bm{x}+\bm{\epsilon},
\end{equation}
where the circuit log-eigenvalues \(\bm{b}\) are related to the gate log-eigenvalues \(\bm{x}\) by the design matrix \(A\).
We assume the error variable is distributed according to a multivariate normal distribution \(\bm{\epsilon}\sim\mathcal{N}_{M}{(0,\Omega^\prime)}\), where \(\Omega^\prime\) is the circuit log-eigenvalue estimator covariance matrix.
As the circuit log-eigenvalues are the logarithm of the circuit eigenvalues, we estimate \(\Omega^\prime\) with the circuit eigenvalue estimator covariance matrix \(\Omega\).
The appropriate first-order Taylor expansion is given elementwise as \(\Omega_{\alpha\beta}^\prime\approx\Omega_{\alpha\beta}/\Lambda_{\alpha}\Lambda_{\beta}\).
This approximation works well in practice and accurately predicts the performance of ACES.

The weighted least squares estimator for the gate log-eigenvalues is
\begin{equation}\label{eq:least-squares-estimator}
    \hat{\bm{x}}={\big(A^\intercal\hat{W}A\big)}^{-1}A^\intercal\hat{W}\bm{b}.
\end{equation}
The weight matrix \(W\) is a diagonal matrix consisting of the inverse circuit log-eigenvalue estimator variances \(W_{\mu\mu}=\Omega_{\mu\mu}^{\prime -1}\).
These variances are estimated with the circuit eigenvalue estimators, so we properly write this as the \emph{estimated weight matrix} \(\hat{W}\).
The gate eigenvalues are at most 1, so the (negative) gate log-eigenvalues are non-negative, and we therefore set negative elements of \(\hat{\bm{x}}\) to 0.
In practice, we reduce weighted least squares to ordinary least squares by left-multiplying \cref{eq:least-squares} by \(\sqrt{\hat{W}}\), allowing us to leverage optimised and numerically stable linear algebra routines for sparse ordinary least squares.

Finally, we estimate the Pauli error probabilities of each gate by taking the Walsh-Hadamard transform of the corresponding gate eigenvalues, following \cref{eq:pauli-probabilities-from-eigenvalues}.
The resulting estimates are not guaranteed to be valid probability distributions, so we project them using the Euclidean distance into the probability simplex with a simple algorithm~\citep{held_validation_1974}, completing our noise estimation procedure.

\subsection{ACES performance analysis}\label{sec:characterising-aces-performance}

We aim to maximise the precision to which we estimate noise given a fixed measurement budget \(S\).
To do this, we analyse the performance of an ACES experimental design \((\mathcal{T},\Gamma)\), which consists of the tuple set \(\mathcal{T}\) and shot weights \(\Gamma\), when characterising a circuit \(\mathcal{C}\) with gate eigenvalues \(\bm{\lambda}\).
The original presentation of ACES suggested the pseudoinverse norm of the design matrix \(A\)~\citep{flammia_averaged_2022}, but we use the circuit eigenvalue estimator covariance matrix described in \cref{eq:eigenvalue-covariance} to precisely calculate the distribution of the gate eigenvalue estimator vector \(\hat{\bm{\lambda}}\) and thereby derive an improved figure of merit.

We calculate the gate log-eigenvalue estimator covariance matrix \(\Sigma^\prime\) by conjugating the circuit log-eigenvalue covariance matrix \(\Omega^\prime\) by the weighted least squares estimation matrix of \cref{eq:least-squares-estimator}, so that
\begin{equation}\label{eq:least-squares-covariance}
    \Sigma^\prime={\big(A^\intercal WA\big)}^{-1}A^\intercal W\Omega^\prime WA{\big(A^\intercal WA\big)}^{-1}.
\end{equation}
We calculate the figure of merit with respect to known values, and so use \(W\) rather than \(\hat{W}\).
As the gate eigenvalues are the exponential of the gate log-eigenvalues, the \emph{gate eigenvalue estimator covariance matrix} \(\Sigma\) follows from \(\Sigma^\prime\) by another first-order Taylor expansion given elementwise by \(\Sigma_{\alpha\beta}\approx\lambda_{\alpha}\lambda_{\beta}\Sigma_{\alpha\beta}^\prime\).
Under our assumptions, the gate eigenvalue estimator residual vector \(\hat{\bm{\lambda}}-\bm{\lambda}\) is distributed according to a multivariate normal distribution \(\mathcal{N}_{N}{(0,\Sigma)}\), as linear least squares is an unbiased estimator.

Now consider the quadratic form \({(\hat{\bm{\lambda}}-\bm{\lambda})}^\intercal(\hat{\bm{\lambda}}-\bm{\lambda})=\lVert\hat{\bm{\lambda}}-\bm{\lambda}\rVert_2^2\), the squared two-norm of the gate eigenvalue residual vector.
It is distributed as a weighted sum of chi-squared random variables with 1 degree of freedom \(\chi_1^2\)---the distribution of the square of a standard normal random variable---where the weights \(\sigma_\nu\) are the eigenvalues of \(\Sigma\), that is
\begin{equation}\label{eq:aces-performance-distribution}
    \lVert\hat{\bm{\lambda}}-\bm{\lambda}\rVert_2^2=\sum_{\nu=1}^{N}{\sigma_\nu y_\nu},\quad y_\nu\sim\chi_1^2.
\end{equation}
This is a generalised chi-squared distribution with mean \(\tr{(\Sigma)}\) and variance \(2\tr{(\Sigma^2)}\).
While it has no known closed-form probability density function~\citep{bodenham_comparison_2016}, this distribution can be calculated numerically~\citep{imhof_computing_1961}.

In particular, we will focus on the normalised root-mean-square (RMS) error \(\sqrt{S^\prime/N}\lVert\hat{\bm{\lambda}}-\bm{\lambda}\rVert_2\) proportional to the square root of the quadratic form.
This quantity averages the error over the number of gate eigenvalues \(N\), and is normalised to remove the dependence on the measurement shots \(S^\prime\) implicitly introduced in \cref{eq:eigenvalue-covariance}.

We seek a figure of merit that is easily computed and accurately predicts the performance of ACES noise characterisation experiments, which will allow us to optimise their experimental design.
To this end, define the \emph{ACES figure of merit} to be the expected normalised RMS error \(\mathcal{F}=\sqrt{S^\prime/N}\expt{\big[\lVert\hat{\bm{\lambda}}-\bm{\lambda}\rVert_2\big]}\).
Smaller figures of merit are better, as the measurement budget required to achieve a fixed estimation accuracy is proportional to the square of the figure of merit.
Note this implies that ACES recovers the usual scaling for additive precision Pauli channel estimation~\citep{fawzi_lower_2023}.

The ACES figure of merit can be approximated by a second-order Taylor expansion as
\begin{equation}\label{eq:aces-figure-of-merit}
    \mathcal{F}(\mathcal{T},\Gamma|\mathcal{C},\bm{\lambda})\approx\sqrt{S^\prime/N}\sqrt{\tr{(\Sigma)}}\left(1-\frac{1}{4}\frac{\tr{(\Sigma^2)}}{\tr{(\Sigma)}^{2}}\right).
\end{equation}
Since we already know the expectation of the square from the quadratic form, the normalised RMS error variance \(\mathcal{V}=S^\prime/N\var{\big[\lVert\hat{\bm{\lambda}}-\bm{\lambda}\rVert_2\big]}\) follows as
\begin{equation}\label{eq:aces-variance}
    \mathcal{V}(\mathcal{T},\Gamma|\mathcal{C},\bm{\lambda})\approx\frac{S^\prime}{2N}\frac{\tr{(\Sigma^2)}}{\tr{(\Sigma)}}\left(1-\frac{1}{8}\frac{\tr{(\Sigma^2)}}{\tr{(\Sigma)}^{2}}\right).
\end{equation}
These approximations are sufficiently accurate in practice that we refer to the approximations and true values interchangeably.
Note that when the shot weights \(\Gamma\) are omitted, we use the default shot weights for \(\mathcal{T}\).

Crucial to a full understanding of ACES is a focus on optimising experimental designs.
Absent this, it is easy to conclude that ACES is only capable of estimating noise to additive precision~\citep{carignan-dugas_error_2023}.
\Cref{apdx:relative-precision} analyses a toy problem to demonstrate that ACES is in general capable of relative precision noise estimation.
However, only certain combinations of gate eigenvalues can be learned to relative precision.
To use the language of cycle error reconstruction~\citep{carignan-dugas_error_2023}, the \emph{orbits} of a gate are the sets of Paulis mapped to each other by the action of the gate, and only the products of gate eigenvalues within each of the orbits of the gate can be learned to relative precision.
For example, consider the orbits of the Hadamard gate \(H\), which are \((X,Z)\) and \((Y)\), and notice that repeating the gate only allows us to amplify the quantities \(\lambda_{H,X}\lambda_{H,Z}\) and \(\lambda_{H,Y}\) and hence estimate them to relative precision.
Empirically, when examining the gate eigenvalue estimator covariance matrix \(\Sigma\) for optimised experimental designs, we find that gate eigenvalue estimators within gate orbits are strongly anticorrelated, such that marginalising over gate orbits obtains a relative precision estimator, though we leave further analysis to future work.

We therefore believe that ACES performs comparably to relative precision Pauli channel estimation techniques in the literature such as cycle error reconstruction (CER)~\citep{carignan-dugas_error_2023}.
In principle, each framework could be modified to obtain the strengths of the other.
For example, our presentation of ACES here focuses on learning a circuit-level noise model, but could be modified to characterise noise that is spatially correlated between gates in a layer, as in CER.
However, this would come at the cost of increasing the number of experiments required to implement the protocol on a quantum device, and our choices in this paper reflect our focus on developing a practical and scalable noise characterisation protocol.

\section{Designing ACES experiments}\label{sec:designing-aces}

Our overview of ACES in \cref{sec:aces-overview} elided the matter of designing ACES noise characterisation experiments, which we now address.
First, we describe a deterministic algorithm that generates the experimental set \(\mathcal{E}_T\) for a tuple \(T\in\mathcal{T}\).
Then, we outline a gradient descent method for optimising the shot weights \(\Gamma\) associated with a tuple set \(\mathcal{T}\).
Lastly, we outline a coordinate descent method for optimising the lengths of tuples, and describe a greedy algorithm that adds random tuples and prunes them from the set, together optimising the tuple set \(\mathcal{T}\).

Our optimisation methods minimise the figure of merit, which is closely related to the trace of the gate eigenvalue estimator covariance matrix.
Consequently, they roughly aim to produce what is known in the theory of optimal experimental designs as an A-optimal design~\citep{fedorov_theory_1972}.
We suspect that this literature offers insights for improving the heuristic algorithms and methods described here.

\subsection{Packing experiment sets}\label{sec:packing-sets}

We now introduce a deterministic algorithm for generating the experiment set \(\mathcal{E}_T\) for a tuple \(T\).
With this algorithm, the tuple set \(\mathcal{T}\) and shot weights \(\Gamma\) fully specify the design of ACES noise characterisation experiments.

Recall from \cref{sec:circuit-eigenvalue-sampling} that the experiment set \(\mathcal{E}_T\) must estimate the circuit eigenvalues \(\Lambda_{T,\bm{a}}\) for all Paulis in the Pauli preparation set \(\bm{a}\in\mathcal{Q}_T\), namely, all Paulis supported on some gate in the rearranged circuit \(\mathcal{C}_T\).
Moreover, if two Paulis \(\bm{a},\bm{a}^\prime\in\mathcal{Q}_T\) are \(T\)-consistent, written \(\bm{a}\tconsistent\bm{a}^\prime\), then we can simultaneously estimate \(\Lambda_{T,\bm{a}}\) and \(\Lambda_{T,\bm{a}^\prime}\) in the same experiment.

\Cref{alg:pack-experiment-set} packs the Pauli preparations \(\mathcal{Q}_T\) into experiments, mutually \(T\)-consistent sets \(E_T\), such that the experiment set \(\mathcal{E}_T=\{E_T\}\) estimates all of the circuit eigenvalues associated with the tuple \(T\).
The \emph{unadded Pauli set} \(\mathcal{U}_T\) is initialised as \(\mathcal{Q}_T\), and tracks the Paulis not yet added to any experiment.
The algorithm adds experiments to the experiment set until it is empty.
The \(T\)-\emph{consistent unadded Pauli set} \(U_T\) and the \(T\)-\emph{consistent Pauli preparation set} \(Q_T\), initialised as \(\mathcal{U}_T\) and \(\mathcal{Q}_T\), respectively, track Paulis that can be added to a specific experiment.
Each time a Pauli \(\bm{a}\) is added to an experiment \(E_T\), we update \(U_T\) and \(Q_T\) by intersecting them with the \(T\)-\emph{consistency set} \(\kappa_T(\bm{a})=\{\bm{a}^\prime\in\mathcal{Q}_T\colon\bm{a}\tconsistent\bm{a}^\prime\}\setminus \{\bm{a}\}\).

Then experiments \(E_T\) are constructed by adding Paulis from \(U_T\) until it is empty, and then adding Paulis from \(Q_T\) until it is also empty.
We sort the Pauli preparations in descending order by the size of their measurement support, and add the first Pauli whose measurement has the maximum overlap with the measurements of the Paulis already in the experiment, as in general measurements can have larger supports than preparations.

\begin{algorithm}
    \caption{Experiment set packing algorithm}
    \label{alg:pack-experiment-set}
    \KwIn{Pauli preparation set \(\mathcal{Q}_T\)}
    \KwOut{Experiment set \(\mathcal{E}_T\)}
    initialise \(\mathcal{E}_T\leftarrow\emptyset\)\;
    sort \(\bm{a}\in\mathcal{Q}_T\) by \(\big|\mspace{-1mu}\supp{(T(\bm{a}))}\mspace{1mu}\big|\) in descending order\;
    initialise \(\mathcal{U}_T\leftarrow\mathcal{Q}_T\)\;
    compute \(\kappa_T(\bm{a})\) for all \(\bm{a}\in\mathcal{Q}_T\)\;
    \tcc{Construct experiments for all Paulis}
    \While{\(\mathcal{U}_T\ne\emptyset\)}{
        initialise \(E_T\leftarrow\emptyset\)\;
        initialise \(U_T\leftarrow\mathcal{U}_T\)\;
        initialise \(Q_T\leftarrow\mathcal{Q}_T\)\;
        \tcc{First add Paulis not yet in any experiment}
        \While{\(U_T\ne\emptyset\)}{
            set \(\bm{a}^\prime\leftarrow\arg{\max\limits_{\bm{a}\in U_T}{\big|\mspace{-1mu}\supp{(T(\bm{a}))}\cap\supp{(E_T)}\mspace{1mu}\big|}}\)\;
            add \(E_T\leftarrow E_T\cup\bm{a}^\prime\)\;
            update \(Q_T\leftarrow Q_T\cap\kappa_T(\bm{a}^\prime)\)\;
            update \(U_T\leftarrow U_t\cap\kappa_T(\bm{a}^\prime)\)\;
            remove \(\mathcal{U}_T\leftarrow\mathcal{U}_T\setminus\bm{a}^\prime\)\;
        }
        \tcc{Then fill the experiment with other Paulis}
        \While{\(Q_T\ne\emptyset\)}{
            set \(\bm{a}^\prime\leftarrow\arg{\max\limits_{\bm{a}\in Q_T}{\big|\mspace{-1mu}\supp{(T(\bm{a}))}\cap\supp{(E_T)}\mspace{1mu}\big|}}\)\;
            add \(E_T\leftarrow E_T\cup\bm{a}^\prime\)\;
            update \(Q_T\leftarrow Q_T\cap\kappa_T(\bm{a}^\prime)\)\;
        }
        add \(\mathcal{E}_T\leftarrow\mathcal{E}_T\cup\{E_T\}\)\;
    }
    \KwRet{\(\mathcal{E}_T\)}
\end{algorithm}

The eigenvalue sampling procedure requires us to prepare all eigenstate sign configurations of each Pauli in the experiment at equal frequency, which is achieved by Pauli frame randomisation.
When practically implementing this protocol, one should choose the number of randomisations for the experiments \(E_T\in\mathcal{E}_T\) of each tuple \(T\) in alignment with its shot weights \(\Gamma_T\), perhaps imposing a minimum.
Then, allocate a constant number of shots to each randomisation, optimally a single shot by leveraging techniques such as in~\citep{fruitwala_hardwareefficient_2024}.

\subsection{Optimising measurement allocation}\label{sec:merit-gradient}

We aim to optimise the shot weights \(\Gamma\), initialised as their default values, of a tuple set \(\mathcal{T}\) with respect to its figure of merit \(\mathcal{F}(\mathcal{T},\Gamma|\mathcal{C},\bm{\lambda})\) when characterising a circuit \(\mathcal{C}\) with gate eigenvalues \(\bm{\lambda}\).
Specifically, we parameterise the shot weights with the \emph{shot log-weights} \(\gamma=\{\gamma_T\}_{T\in\mathcal{T}}\) as \(\Gamma_T=\exp{(-\gamma_T)}/\sum_{U\in\mathcal{T}}{\exp{(-\gamma_U)}}\), and perform gradient descent on the shot log-weights \(\gamma\).

Nesterov momentum augments gradient descent with a velocity that tracks gradients across update steps~\citep{nesterov_method_1983, sutskever_importance_2013}.
The learning rate \(\eta>0\) controls the size of the contribution of the gradient to the velocity at each step, whereas the momentum coefficient \(\mu\in[0,1]\) controls the decay in the velocity across update steps.
Gradient descent corresponds to \(\mu=0\).
Writing the shot log-weights as a vector \(\bm{\gamma}\), the update step is given by
\begin{align}
    \bm{v}^{(s+1)}&=\mu\bm{v}^{(s)}-\eta\frac{\partial\mathcal{F}}{\partial\bm{\gamma}}\big(\mathcal{T},\bm{\gamma}^{(s)}+\mu\bm{v}^{(s)}\big),\\
    \bm{\gamma}^{(s+1)}&=\bm{\gamma}^{(s)}+\bm{v}^{(s+1)}.
\end{align}
In practice, we find it helpful to revert update steps that worsen the figure of merit and then zero the velocity.
If this happens multiple times in quick succession, we also reduce the learning rate \(\eta\) by a factor \(\eta_r\).
We usually choose parameter values \(\eta=10^{3/4}\), \(\mu=0.99\), and \(\eta_r=10^{1/4}\).

Gradient calculations are optimised with the analytic expressions in \cref{apdx:merit-differentiation}.

\subsection{Optimising tuple sets}\label{sec:tuple-optimisation}

When introducing tuple sets in \cref{sec:aces-fundamentals}, we only described how to construct the basic tuple set \(\mathcal{T}_{\mathcal{I}}\), which describes the most basic ACES noise characterisation experiment.
We now outline two algorithms that allow us to optimise tuple sets \(\mathcal{T}\) by their figure of merit.

Elfving's theorem characterises optimal designs for linear regression problems~\citep{elfving_optimum_1952,studden_elfvings_1971}, which use a minimal number of maximally spaced points.
In the context of ACES, shallow tuples produce large circuit eigenvalues near \(1\), whereas deep tuples with a large number of layers produce smaller eigenvalues.
We therefore augment the basic tuple set, containing shallow tuples, with the \emph{repeated tuple set} \(\mathcal{T}_{\mathrm{rep}}\), whose construction we describe in \cref{apdx:tuple-generation}.
The tuples in \(\mathcal{T}_{\mathrm{rep}}\), which resemble the germs of gate set tomography~\citep{nielsen_gate_2021}, are parameterised by repetition numbers describing how many times the elements of the tuple are repeated.
While we could optimise our choice of repeated tuples, we do not do so here.

We optimise these repetition numbers with a heuristic gradient-free coordinate descent algorithm~\citep{wright_coordinate_2015}, which cycles through the repetition numbers and steps them towards smaller figures of merit.
The algorithm grows the step size exponentially when repeatedly stepping in the same direction, and optimises the shot weights \(\Gamma\) with the gradient descent routine in \cref{sec:merit-gradient} before evaluating the figure of merit.
In practice, it appears most performant to choose repeated tuples that, up to a Pauli correction, implement an involution, so that repeating the tuple twice implements the identity, and use only odd repetition numbers.

After optimising the repetition numbers of the deep tuples, we optimise the shallow tuples with \cref{alg:optimise-tuple-set}, a greedy algorithm that resembles the experimental design optimisation algorithm of~\citep{mitchell_algorithm_1974}.
It performs a number of \emph{excursions} \(n_{\mathrm{ex}}\), each of which greedily grows and shrinks the tuple set.
To grow the tuple set, an excursion generates random tuples according to a probability distribution over tuples \(\mathcal{P}\) and greedily adds them to the tuple set according to the figure of merit.
It does this until the tuple set contains \(l_{\mathrm{set}}+l_{\mathrm{ex}}\) tuples, where \(l_{\mathrm{set}}\) is the desired tuple set size and \(l_{\mathrm{ex}}\) is the excursion length, by trialling up to \(f_{\mathrm{trial}}\) times more tuples than it is trying to add.
Then to shrink the tuple set, it greedily prunes tuples until either it can no longer remove a tuple without increasing the figure of merit, or the tuple set contains \(l_{\mathrm{set}}\) tuples.

Although \cref{alg:optimise-tuple-set} could in principle remove deep tuples, or add tuples of intermediate depth, it tends not to do so in practice.
We usually choose \(n_{\mathrm{ex}}=3\), \(l_{\mathrm{ex}}=10\), \(l_{\mathrm{set}}=5|\mathcal{I}|\), and \(f_{\mathrm{trial}}=20\), and describe \(\mathcal{P}\) in \cref{apdx:tuple-generation}.

\begin{algorithm}
    \caption{Tuple set optimisation algorithm}
    \label{alg:optimise-tuple-set}
    \KwIn{tuple set \(\mathcal{T}\), circuit \(\mathcal{C}\), gate eigenvalues \(\bm{\lambda}\)}
    \KwOut{tuple set \(\mathcal{T}\)}
    \SetKwInOut{KwParam}{Parameters}
    \KwParam{Excursion number \(n_{\mathrm{ex}}\) and length \(l_{\mathrm{ex}}\), tuple set size \(l_{\mathrm{set}}\), trial factor \(f_{\mathrm{trial}}\), tuple probability distribution \(\mathcal{P}\)}
    \For{\(\mathrm{ex}\) \In \([n_{\mathrm{ex}}]\)}{
    \tcc{Greedily add tuples to the set}
    initialise \(t\leftarrow f_{\mathrm{trial}}\big(l_{\mathrm{set}}+l_{\mathrm{ex}}-|\mathcal{T}|\big)\)\;
    \While{\(|\mathcal{T}|<l_{\mathrm{set}}+l_{\mathrm{ex}}\) \And \(t>0\)
    }{
    sample \(T\sim\mathcal{P}\)\;
    \If{\(\mathcal{F}(\mathcal{T}\cup T|\mathcal{C},\bm{\lambda})<\mathcal{F}(\mathcal{T}|\mathcal{C},\bm{\lambda})\)}{
        append \(\mathcal{T}\leftarrow \mathcal{T}\cup T\)\;
    }
    decrement \(t\leftarrow t-1\)\;
    }
    \tcc{Greedily remove tuples from the set}
    \While{\(\mathcal{F}(\mathcal{T}\setminus T^\prime|\mathcal{C},\bm{\lambda})<\mathcal{F}(\mathcal{T}|\mathcal{C},\bm{\lambda})\) \Or \(|\mathcal{T}|>l_{\mathrm{set}}\)}{
    set \(T^\prime\leftarrow\arg\min_{T\in\mathcal{T}}{\mathcal{F}(\mathcal{T}\setminus T|\mathcal{C},\bm{\lambda})}\)\;
    remove \(\mathcal{T}\leftarrow \mathcal{T}\setminus T^\prime\)\;
    }
    }
    \KwRet{\(\mathcal{T}\)}
\end{algorithm}

After optimising the tuple set \(\mathcal{T}\) with these algorithms, we might in practice add repeated tuples of an intermediate depth to verify the expected exponential decay in the circuit eigenvalues as a function of circuit depth at the cost of performance.
Finally, we optimise the shot weights \(\Gamma\) to obtain an optimised experimental design \((\mathcal{T},\Gamma)\).

\section{Numerical results}\label{sec:numerical-results}

\begin{figure}[b]
    \centering
    \includegraphics[width=\columnwidth]{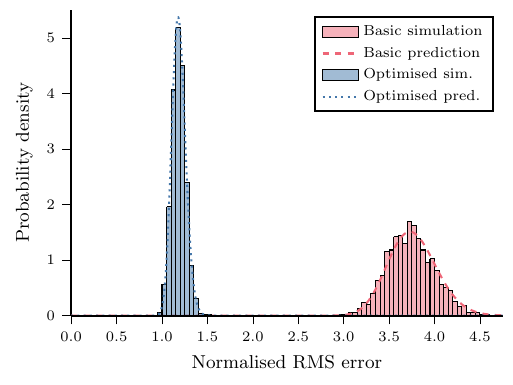}
    \vspace{-2em}
    \caption{Simulated and predicted performance of basic and optimised ACES experimental designs.
    Simulated data show the distribution of the normalised RMS error between the estimated and true gate eigenvalues when characterising a fixed-seed random instance of log-normal Pauli noise for the syndrome extraction circuit of a distance-\(3\) surface code.
    The data are \(1000\) trials of characterising a fixed-seed random instance of log-normal Pauli noise with a measurement budget \(S=10^8\) and align with predicted performance distributions based on \cref{eq:aces-performance-distribution}.
    The optimised experimental design substantially outperforms the basic experimental design.}
    \label{fig:aces-performance}
\end{figure}

\begin{figure*}[t]
    \centering
    \includegraphics[width=\textwidth]{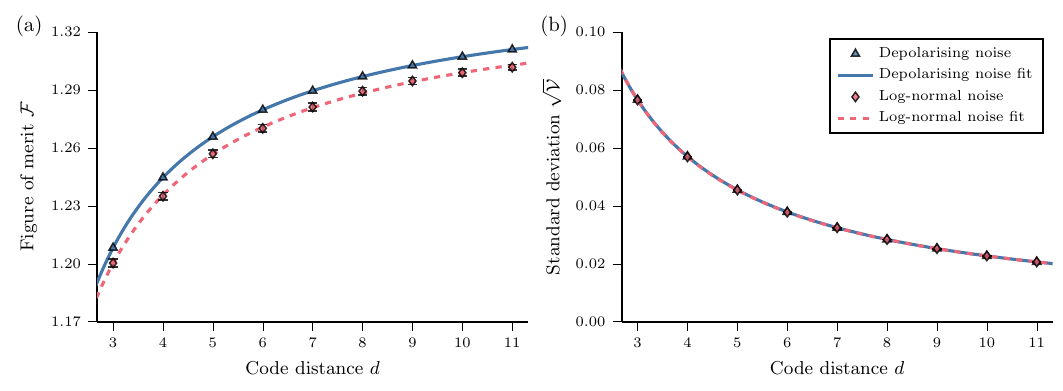}
    \vspace{-2em}
    \caption{Performance scaling of ACES noise characterisation as a function of the surface code distance \(d\).
    (a) Figure of merit \(\mathcal{F}\), the expected normalised RMS error, and (b) the normalised RMS error standard deviation \(\sqrt{\mathcal{V}}\), of the optimised experimental design for the syndrome extraction circuit.
    The data points have no error for depolarising noise, whereas for log-normal Pauli noise, they report the mean estimated by sampling from the distribution over noise models, with error bars indicating two standard deviations.
    The functional forms, based on \cref{eq:aces-figure-of-merit,eq:aces-variance} and discussed in the text, fit the data precisely, with relative errors of under \(10^{-8}\) for each depolarising noise data point.}
    \label{fig:merit-distance-scaling}
\end{figure*}

We now present numerical results demonstrating that our methods are scalable and capable of performant noise characterisation of surface code syndrome extraction circuit.
These results were produced on a 2021 M1 Max Macbook Pro with 32 GB of RAM with Julia~\citep{bezanson_julia_2017}, and stabiliser circuits were stimulated with Stim~\citep{gidney_stim_2021}.
We release all of our code as the Julia package QuantumACES~\citep{hockings_quantumacesjl_2025}.

We test our methods on a distribution over noise models that we call \emph{log-normal Pauli noise}, where each Pauli error probability associated with each gate is independently log-normally distributed, as this resembles the noise observed in the quantum device featured in the surface code experiment~\citep{acharya_suppressing_2023}.
We choose physically-relevant average single-qubit gate, two-qubit gate, and measurement error rates of \(r_1=0.075\%\), \(r_2=0.5\%\), and \(r_m=2\%\), respectively, and supply the full details of the noise model, including layer times, in \cref{apdx:experimental-noise}.
We will test our noise characterisation methods on the seed-\(0\) random instance of this noise model, and optimise our experimental designs for depolarising noise with the same error rates, where under depolarising noise gates have an equal probability of each Pauli error.

We first optimise an experimental design for depolarising noise on the syndrome extraction circuit of a distance-\(3\) surface code on \(17\) qubits, using the procedures outlined in \cref{sec:designing-aces}.
This depolarising noise model functions only as an optimisation target and is constructed from estimates of the average single-qubit gate, two-qubit gate, and measurement error rates, in this case as detailed above for log-normal Pauli noise.
The random tuples added by \cref{alg:optimise-tuple-set} are shallow, with depth at most \(4\), and the shot weights allocate the majority of the measurement budget to individually measuring the controlled-\(Z\) gate layers.
We present the complete experimental design in \cref{apdx:add-results}.

\begin{figure}[b]
    \centering
    \includegraphics[width=\columnwidth]{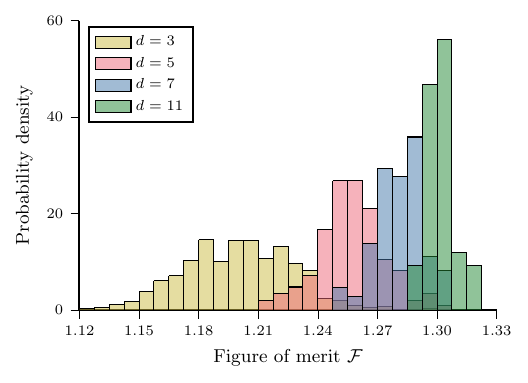}
    \vspace{-2em}
    \caption{Variation in the performance of ACES noise characterisation across random instances of log-normal Pauli noise at a range of surface code distances \(d\).
    The histograms indicate the distribution of the figure of merit \(\mathcal{F}\), the expected normalised RMS error, of the optimised experimental design for the syndrome extraction circuit.
    Random instances of log-normal Pauli noise have increasingly similar figures of merit with increasing surface code distance \(d\).}
    \label{fig:log-normal-merit-distribution}
\end{figure}

\begin{figure*}[t!]
    \centering
    \includegraphics[width=\textwidth]{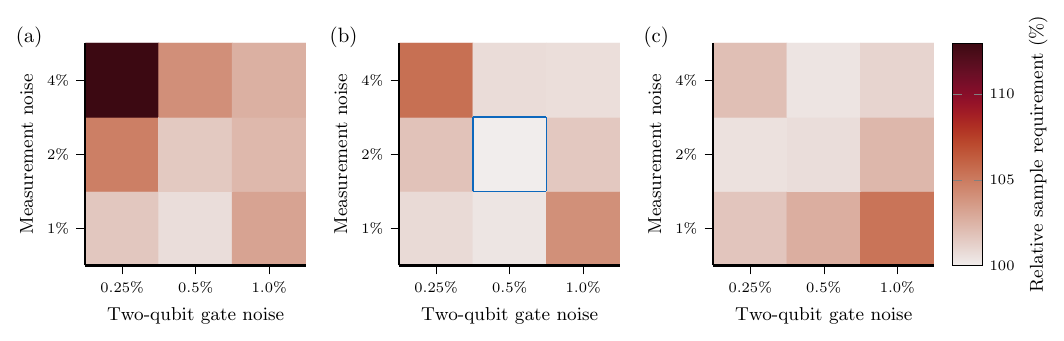}
    \vspace{-2em}
    \caption{Performance of ACES experimental designs optimised for inaccurate average error rates.
    The experimental designs were optimised for depolarising noise with a range of error rates, with single-qubit gate error rates (a) \(r_1=0.0375\%\), (b) \(r_1=0.075\%\), and (c) \(r_1=0.15\%\), and two-qubit gate and measurement error rates indicated in the heatmaps.
    Experimental designs were evaluated for the syndrome extraction circuit of a distance-\(3\) surface code against \(400\) random instances of log-normal Pauli noise, whose average error rates are indicated by the blue square.
    Heatmap colour indicates the expected number of samples required to achieve a fixed estimation accuracy, expressed relative to the best-performing design optimised at the correct average error rates.
    Only a single experimental design requires over \(10\%\) more samples to achieve the same accuracy as the best-performing design.}
    \label{fig:merit-heatmap}
\end{figure*}

Then we perform ACES noise characterisation on a fixed-seed random instance of log-normal Pauli noise, comparing the basic and optimised experimental designs.
\Cref{fig:aces-performance} shows the distribution of the normalised RMS error over \(1000\) simulated trials of ACES with a measurement budget \(S=10^8\) for both designs, as well as the predicted normalised RMS error distributions.
The predicted distributions closely align with the simulated data, validating our ability to predict the performance of ACES noise characterisation experiments, and demonstrating that the optimised design substantially outperforms the basic design.
For this random instance of log-normal Pauli noise, the basic design has a figure of merit, or expected normalised RMS error, that is larger than the optimised design by a factor \(3.17\), implying that the sample efficiency of the optimised design is a factor \(10.1\) better than the basic design.

Next, we examine how the performance of this optimised experimental design for the syndrome extraction circuit varies as a function of the surface code distance.
\Cref{fig:merit-distance-scaling} depicts the figure of merit \(\mathcal{F}\), the expected normalised RMS error, and the normalised RMS error standard deviation \(\sqrt{\mathcal{V}}\) as functions of the surface code distance \(d\).
The exact quantities are reported for depolarising noise, whereas for log-normal Pauli noise, we report estimates of the mean across random samples from the distribution over noise models.

Empirically, we find that for depolarising noise, the normalised trace of the gate eigenvalue estimator covariance matrix \(\tr{(\Sigma)}/S^\prime\), and of its square \(\tr{(\Sigma^2)}/S^{\prime 2}\), are precisely fit as quadratic functions of \(d\).
Model selection using the Akaike information criterion corrected for small samples~\citep{hurvich_regression_1989} prefers this quadratic model over other polynomial models.
Similarly, the number of gate eigenvalues is exactly described by a quadratic with integer coefficients, \(N(d)=84d^2-36d-24\).
Substituting these three quadratics into \cref{eq:aces-figure-of-merit,eq:aces-variance} yields functional forms that precisely describe the scaling of \(\mathcal{F}\) and \(\mathcal{\sqrt{V}}\) with \(d\), which are also depicted in \cref{fig:merit-distance-scaling}.
For depolarising noise, we fit the normalised traces of the gate eigenvalue estimator covariance matrix, obtaining functional forms for \(\mathcal{F}\) and \(\mathcal{\sqrt{V}}\) with relative errors of less than \(10^{-8}\) for each data point.
By contrast, for log-normal Pauli noise, we directly and simultaneously fit the functional forms for \(\mathcal{F}\) and \(\mathcal{\sqrt{V}}\).
Although the data represent estimates, we nevertheless obtain relative errors of less than \(10^{-3}\) for each data point.

\begin{figure}[b]
    \centering
    \includegraphics[width=\columnwidth]{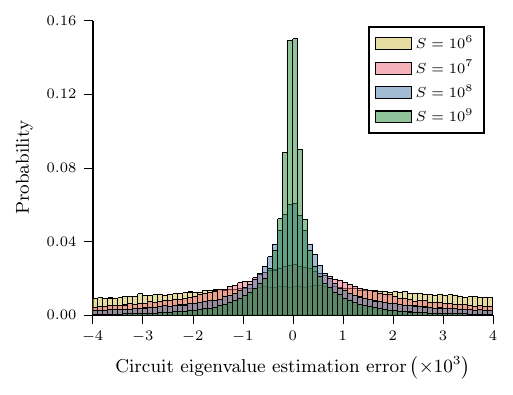}
    \vspace{-2em}
    \caption{Performance of ACES noise characterisation with the optimised experimental design for the syndrome extraction circuit of a distance-\(25\) surface code.
    The histograms indicate distributions of the circuit eigenvalue estimation error, the difference between the estimated and true circuit eigenvalues.
    The data are for a fixed-seed random instance of log-normal Pauli noise over a range of measurement budgets \(S\).
    Circuit eigenvalues are estimated to greater precision with increasing \(S\).}
    \label{fig:aces-eigenvalue-histogram}
\end{figure}

\begin{figure*}[t!]
    \centering
    \includegraphics[width=\textwidth]{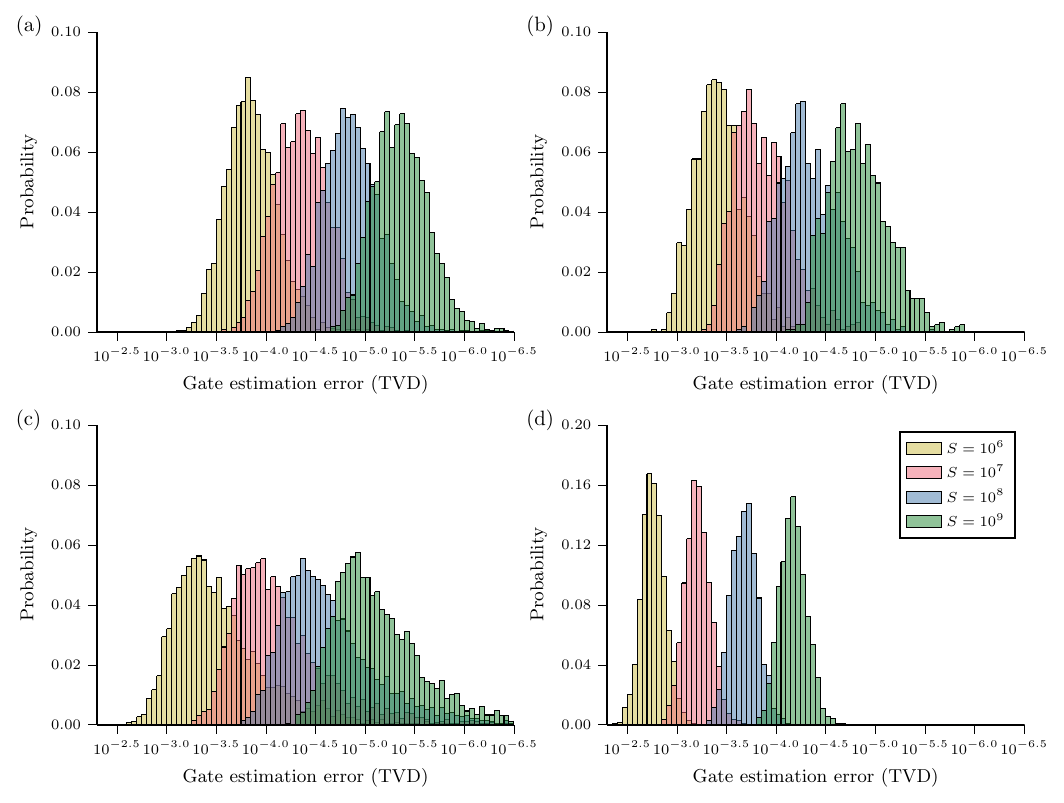}
    \vspace{-2em}
    \caption{Performance of ACES noise characterisation with the optimised experimental design for the syndrome extraction circuit of a distance-\(25\) surface code.
    The histograms indicate distributions of the gate estimation error, the total variation distance (TVD) between the estimated and true error probability distributions for the gate.
    The data are for a fixed-seed random instance of log-normal Pauli noise over a range of measurement budgets \(S\).
    The estimation error distributions are shown across the range of gate types appearing in the circuit: (a) dynamical decoupling \(X\) gates and padded identity gates, or Pauli gates; (b) Hadamard gates; (c) measurements; and (d) controlled-\(Z\) gates.
    Across all gate types, the estimation error distributions consistently shift roughly by a constant factor of \(1/\sqrt{10}\) when the measurement budget \(S\) is increased by a factor of \(10\), demonstrating the anticipated sample efficiency.}
    \label{fig:aces-tvd-histograms}
\end{figure*}

Importantly, the functional forms derived by substituting the quadratics into \cref{eq:aces-figure-of-merit,eq:aces-variance} entail that in the limit of large \(d\), the figure of merit \(\mathcal{F}\), or expected normalised RMS error, approaches a constant value.
Also, the optimised experimental design requires \(261\) experiments, before Pauli frame randomisation, to estimate all of the circuit eigenvalues at all tested code distances \(d\), which range from \(d=3\) to \(d=25\).
This demonstrates that ACES is capable of estimating noise in surface code syndrome extraction circuits to a precision that is asymptotically independent of the number of qubits \(n\) in the code, using a number of experiments that is also asymptotically independent of \(n\).

Notably, the optimised experimental design performs better on average for log-normal Pauli noise, despite being optimised for depolarising noise.
Moreover, \cref{fig:log-normal-merit-distribution} shows histograms of the figure of merit \(\mathcal{F}\) across random instances of log-normal Pauli noise at a range of surface code distances \(d\).
It shows that random samples of log-normal Pauli noise have increasingly similar figures of merit as the surface code distance \(d\) increases.
Together, these suggest that the performance of the optimised experimental design is not substantially harmed by small changes in the noise model that preserve average error rates.
The effects of randomness in the noise model on the performance of ACES also appear to average out with increasing system size.

We also examine experimental designs optimised at different error rates to determine the robustness of their performance.
Specifically, we optimise for half the error rate, the error rate, and double the error rate, on single-qubit gates, two-qubit gates, and measurements, yielding \(27\) different experimental designs.
\Cref{fig:merit-heatmap} shows the expected number of samples required to achieve a fixed estimation accuracy, evaluated over \(400\) random instances of log-normal Pauli noise for the syndrome extraction circuit of a distance-\(3\) surface code.
These values are expressed relative to the original experimental design optimised for the appropriate average error rates, which performs best.

Only \(3\) of the \(27\) experimental designs require over \(5\%\) more samples than the most performant experimental design to achieve the same expected accuracy, and the only one to require greater than \(10\%\) more samples specifically requires \(13.00\pm 0.10\%\).
This corresponds to having a mean figure of merit of \(1.2757\pm 0.0016\), whereas the mean figure of merit of the most performant design is \(1.2001\pm 0.0014\).
The expected relative number of samples required to achieve the same accuracy is the square of this ratio, and is estimated precisely due to substantial covariance in the figure of merit between designs for each instance of log-normal Pauli noise.
Overall, this demonstrates that the performance of optimised experimental designs is robust to being optimised for depolarising noise with inappropriate average error rates.

Finally, we use our optimised experimental design to characterise a fixed-seed random instance of log-normal Pauli noise on the syndrome extraction circuit of a distance \(d=25\) surface code with \(n=1249\) qubits.
The optimised experimental design features \(31\) tuples which together estimate \(267,357\) circuit eigenvalues over \(261\) experiments, before Pauli frame randomisation, in order to estimate the \(51,576\) gate eigenvalues.
Optimising the experimental design at \(d=3\), generating it at \(d=25\), and then simulating ACES noise characterisation experiments for measurement budgets \(S\in\{10^6,10^7,10^8,10^9\}\) together took under \(5\) hours, with roughly half of that time being dedicated to stabiliser circuit simulations with Stim.
The fits in \cref{fig:merit-distance-scaling} predict an average figure of merit \(1.3245\) and a standard deviation \(0.0091\), across random instances of log-normal Pauli noise.
By comparison, for the simulated noise characterisation of a specific random instance of log-normal Pauli noise, the normalised RMS errors were \(\{1.4480, 1.3243, 1.3303, 1.3245\}\) across measurement budgets.
Excepting the smallest measurement budget \(S=10^6\), these results are highly consistent with our performance predictions.

The noise estimation procedure begins by estimating the circuit eigenvalues from raw measurement data.
\Cref{fig:aces-eigenvalue-histogram} depicts histograms of the circuit eigenvalue estimation error, the difference between the estimated and true circuit eigenvalues, across the range of measurement budgets.
As the measurement budget increases, the distributions of the circuit eigenvalue estimation error narrow, straightforwardly improving the accuracy of the circuit eigenvalue estimates.

\begin{table}
    \centering
    \vspace{-1.25em}
    \caption{Median of the gate estimation error distributions shown in \cref{fig:aces-tvd-histograms}.
    Results are shown for the optimised experimental design over a range of measurement budgets \(S\), and across the range of gate types appearing in the circuit: dynamical decoupling \(X\) gates and padded identity gates, or Pauli gates; Hadamard gates; measurements; and controlled-\(Z\) gates.
    Sequential entries differ by roughly \(0.5\), confirming the anticipated \(1/\sqrt{S}\) scaling of the estimation error.}
    \vspace{0.25em}
    \def\arraystretch{1.4}
    \begin{tabular}{|c|c|c|c|c|}
        \hline
        \multirow[c]{2}{5.75em}{\centering Measurement budget} & \multicolumn{4}{c|}{\(-\log_{10}\) median gate estimation error (TVD)} \\
        \cline{2-5}
                 & Pauli & Hadamard & Measurement & Controlled-\(Z\) \\
        \hline
        \(10^6\) & 3.841 & 3.415    & 3.437       & 2.761            \\
        \(10^7\) & 4.353 & 3.829    & 4.045       & 3.205            \\
        \(10^8\) & 4.850 & 4.313    & 4.548       & 3.680            \\
        \(10^9\) & 5.348 & 4.805    & 5.051       & 4.171            \\
        \hline
    \end{tabular}
    \label{tab:median-tvd}
\end{table}

Ultimately, the protocol estimates the Pauli error probability distributions of all gates and measurements appearing in the circuit.
We measure the gate estimation error with the total variation distance (TVD) between the estimated and true Pauli error probability distribution, a principled measure for probability distributions and equivalent to the diamond norm between Pauli channels~\citep{fawzi_lower_2023}.
\Cref{fig:aces-tvd-histograms} depicts histograms of the gate estimation error across measurement budgets and the different types of gate appearing in the circuit, namely: dynamical decoupling \(X\) gates and padded identity gates, which are Pauli gates; Hadamard gates; measurements; and controlled-\(Z\) gates.
The Pauli gates tend to be estimated to a higher precision than the Hadamard gates, despite both being single qubit gates.
This is potentially related to the fact that Pauli gates commute with other Pauli gates, up to sign, a setting in which it is easier to achieve relative precision noise estimation, as discussed in \cref{apdx:relative-precision}.
By contrast, the two-qubit controlled-\(Z\) gates are least accurately estimated, accounted for in part by their Pauli error probability distribution being over \(16\) errors, compared to \(4\) for the single-qubit gates and just \(2\) for measurements.

The gate estimation errors improve roughly by a constant factor of \(\sqrt{10}\) for each factor of \(10\) increase in the measurement budget \(S\), demonstrating the expected \(1/\sqrt{S}\) sample efficiency.
This is clear visually and quantified in \cref{tab:median-tvd}, which shows the median gate estimation error across gate types and measurement budgets.
The Hadamard and controlled-\(Z\) gate estimates outperform the trend, particularly for the smallest measurement budget \(S=10^6\).
This is because large estimation errors for small values of \(S\) are improved by ensuring parameter estimates are within bounds.
Specifically, gate eigenvalues greater than \(1\) are set to \(1\), and estimated probability distributions are projected into the simplex.
For the data in \cref{tab:median-tvd}, at \(S=10^6\), \(1211\) gate eigenvalues are set to \(1\), while at \(S=10^7\), only \(51\) gate eigenvalues are set to \(1\).

By contrast, referring again to \cref{tab:median-tvd}, we see that measurement error estimation underperforms the trend at \(S=10^6\).
This appears to be the cause of the outlying poor normalised RMS error \(1.4480\) at this measurement budget.
Our performance prediction calculations implicitly assume that we are in the limit of collecting a large number of samples, so poor prediction at the smallest measurement budget is not surprising.
Nevertheless, performance is highly consistent with our predictions at all larger measurement budgets.

We have demonstrated scalable and performant noise characterisation of the syndrome extraction circuits of very large surface codes in a physically-relevant parameter regime.
We are able to precisely predict the performance of ACES noise characterisation experiments, including as functions of the surface code distance.
Optimising experimental designs for our figure of merit robustly improves performance, even for a different noise model or average error rates.
We supply further numerical results in \cref{apdx:add-results}, and reproduce them for the unrotated surface code in \cref{apdx:unrotated-surface-codes}.

\section{Conclusions}\label{sec:conclusions}

In this paper, we have described and numerically validated a scalable noise characterisation protocol based on ACES.
Our protocol generates optimised experimental designs, leveraging our precise analysis of the performance of ACES noise characterisation experiments.
We show in numerical simulations of a distance-\(25\) surface code with \(1249\) qubits that we can achieve characterisation of circuit-level Pauli noise with predictable performance.
We believe the methods outlined here could be extended to all components of fault-tolerant quantum computation with topological quantum codes.

There are a large number of potential directions for future work.
A rigorous and general theoretical understanding of tuple sets and their performance characteristics would be clarifying, and would likely suggest improved optimisation algorithms.
In particular, our empirical ability to precisely predict the performance scaling of ACES noise characterisation of syndrome extraction circuits as a function of the surface code distance suggests underlying structure that we have not uncovered here.
We may be able to improve the performance of ACES by drawing more heavily on the theory of optimal experimental designs~\citep{fedorov_theory_1972}, as well as on experimental design techniques used in other noise characterisation methods.
Salient additions include optimising our choice of repeated tuples and removing measurements of circuit eigenvalues that do not improve performance, analogous to germ selection and fiducial-pair reduction in gate set tomography~\citep{nielsen_gate_2021}.
A detailed investigation of the relative precision estimation capabilities of ACES would require marginalising noise estimates across gate Pauli orbits, as in cycle error reconstruction~\citep{carignan-dugas_error_2023}, and may require projecting noise estimates into the probability simplex using the Mahalanobis distance~\citep{gelman_causal_2006}, rather than the Euclidean distance, to retain covariance matrix information.

A natural next step would be practically implementing our protocol to characterise noise in a quantum device operating the syndrome extraction circuit of a topological quantum code.
Operating a quantum error correcting code entails mid-circuit measurement and reset as part of syndrome extraction, and although we do not characterise the associated noise here, we believe ACES can straightforwardly be modified to characterise the associated noise, following methods such as those in~\citep{zhang_generalized_2025, hines_pauli_2025}.
The resulting noise estimates could then be used for error mitigation, numerical simulations, or supplied to a noise-aware decoder to determine the extent to which detailed noise characterisation can improve code performance.

\begin{acknowledgments}
We thank Steven T. Flammia for discussions.
ETH is supported by an Australian Government Research Training Program Scholarship.
RH is supported by the Sydney Quantum Academy.
This work was supported by the Australian Research Council Centre of Excellence for Engineered Quantum Systems (CE170100009) and the U.S. Army Research Office (W911NF-21-1-0001).
\end{acknowledgments}

\bibliography{scalable-aces}

\clearpage

\appendix

\section{ACES estimates marginal Pauli error probabilities}\label{apdx:estimate-marginal}

When presenting ACES in \cref{sec:aces-fundamentals}, we assumed that each layer \(\mathcal{C}_i\) of a circuit \(\mathcal{C}\) acting on \(n\) qubits is associated with an \(n\)-qubit Pauli noise channel \(\mathcal{E}_i\).
As it is generally intractable to learn all \(4^n\) eigenvalues of \(n\)-qubit Pauli channels, we focus on learning a circuit-level noise model.
Each layer \(\mathcal{C}_i\) is decomposed into gates \(\mathcal{G}_{ij}\) supported on mutually disjoint sets of qubits, and we learn Pauli channels \(\mathcal{E}_{ij}\) associated with the gates \(\mathcal{G}_{ij}\).

When estimating the eigenvalues of the gate Pauli channel \(\mathcal{E}_{ij}\), we in fact estimate a subset of the eigenvalues of the layer Pauli channel \(\mathcal{E}_i\), namely those eigenvalues corresponding to Paulis supported on the gate \(\mathcal{G}_{ij}\).
It is a consequence of Lemma 4, and by extension Equations 36 and 37, of~\citep{flammia_efficient_2020}, that the corresponding Pauli error probabilities of the gate channel \(\mathcal{E}_{ij}\) are the marginal of the Pauli error probability distribution of the layer channel \(\mathcal{E}_i\) onto the support of the gate \(\mathcal{G}_{ij}\).
Hence when ACES estimates the noise channel \(\mathcal{E}_{ij}\) of the gate \(\mathcal{G}_{ij}\) in the context of the layer \(\mathcal{C}_i\), it captures the averaged effect on that particular gate of the spatial correlations in the full \(n\)-qubit noise channel.

We now prove this claim directly, using the Pauli transfer matrix and superoperator formalism described in, for example,~\citep{greenbaum_introduction_2015}, but omitting much of the formalism of~\citep{flammia_efficient_2020}.
Write \(\bm{\lambda}\) to refer to the vector of eigenvalues of \(\mathcal{E}_i\), namely \(\lambda_{\bm{a}}\) for \(\bm{a}\in\mathbf{P}^n\), and similarly write the vector of Pauli error probabilities as \(\bm{p}\).
As explained in \cref{eq:pauli-eigenvalues-from-probabilities,eq:pauli-probabilities-from-eigenvalues}, these are related as \(\bm{\lambda}=W\bm{p}\) by a Walsh-Hadamard transform matrix \(W\) ordered by the symplectic form \(\omega\), given as
\begin{equation}\label{eq:walsh-hadamard}
    W=\sum_{\bm{a},\bm{b}\in\mathbf{P}^n}{{(-1)}^{\omega(\bm{a},\bm{b})}|\bm{a}\rangle\rangle\langle\langle\bm{b}|}.
\end{equation}
This matrix is its own inverse up to a constant factor of \(4^{-n}\), as \(W^2=4^n I\), allowing us to easily estimate \(\bm{p}\) from \(\bm{\lambda}\).

Consider a subgroup \(A\subseteq\mathbf{P}^n\), and write \(\bm{\lambda}_A\) to refer to the eigenvalues \(\lambda_{\bm{a}}\) for \(\bm{a}\in A\).
These are related as \(\bm{\lambda}_A=\Pi_A\bm{\lambda}\) by the projector onto \(A\), which is simply
\begin{equation}\label{eq:projector}
    \Pi_A=\sum_{\bm{a}\in A}{|\bm{a}\rangle\rangle\langle\langle\bm{a}|}.
\end{equation}
We seek to determine the properties of the Pauli error probability distribution estimated from \(\bm{\lambda}_A\).
To avoid introducing additional formalism from~\citep{flammia_efficient_2020}, we specialise our analysis to the subgroup \(A=\Pauli{(\mathcal{G}_{ij})}\) of Paulis supported on some gate \(\mathcal{G}_{ij}\) occurring in the layer \(\mathcal{C}_i\).
This does not place any restrictions on the support of \(\mathcal{G}_{ij}\), which we write as \(S_{ij}=\supp{(\mathcal{G}_{ij})}\subseteq[n]\), an arbitrary subset of the qubits.
In this case, the eigenvalues \(\bm{\lambda}_A\) are precisely the eigenvalues of \(\mathcal{E}_i\) we measure in the process of estimating \(\mathcal{E}_{ij}\).

Instead denoting \(A=\Pauli{(\mathcal{G}_{ij})}=\Pauli{(S_{ij})}\) allows us to write \(A^C=\Pauli{(S_{ij}^C)}=\Pauli{([n]\setminus S_{ij})}\) as the subgroup of Paulis supported on the set complement \(S_{ij}^C=[n]\setminus S_{ij}\).
Then, we can clearly write any Pauli \(\bm{a}\in\mathbf{P}^n\) as a sum \(\bm{a}=\bm{b}+\bm{c}\) for some \(\bm{b}\in A\) and \(\bm{c}\in A^C\).
Indeed, for any \(\bm{b}\in A\) and \(\bm{c}\in A^C\), \(\omega(\bm{b},\bm{c})=0\) as Paulis supported on mutually disjoint sets of qubits trivially commute.
In the language of~\citep{flammia_efficient_2020}, this is because \(A^C\) is the commutant of \(A\) and \(A\) is its own anticommutant.

This allows us to rewrite \cref{eq:pauli-eigenvalues-from-probabilities} for eigenvalues \(\lambda_{\bm{a}}\) such that \(\bm{a}\in A\), which yields
\begin{align}\begin{split}
        \lambda_{\bm{a}}&=\sum_{\bm{b}\in A}{\sum_{\bm{c}\in A^C}{{(-1)}^{\omega(\bm{a},\bm{b}+\bm{c})}p_{\bm{b}+\bm{c}}}}\\
        &=\sum_{\bm{b}\in A}{{(-1)}^{\omega(\bm{a},\bm{b})}\left(\sum_{\bm{c}\in A^C}{p_{\bm{b}+\bm{c}}}\right)}.
    \end{split}\end{align}
That is, the eigenvalues \(\bm{\lambda}_A\) depend only on the distribution \(\bm{p}\) marginalised over the Paulis in \(A^C\) onto the Paulis in \(A\).
We can write this as \(\bm{p}_A\), being careful to distinguish the meaning from \(\bm{\lambda}_A\), which refers to the eigenvalues for the Paulis in \(A\).

Indeed, the eigenvalues \(\bm{\lambda}_A\) are related to the marginal distribution \(\bm{p}_A\) by the projection of the Walsh-Hadamard transform onto \(A\), namely \(W_A=\Pi_A W\Pi_A\).
Calculating, we see that
\begin{align}\begin{split}
        W_A^2&=\sum_{\bm{a},\bm{a}^\prime,\bm{b},\bm{b}^\prime\in A}{{(-1)}^{\omega(\bm{a},\bm{b})+\omega(\bm{a}^\prime,\bm{b}^\prime)}|\bm{b}^\prime\rangle\rangle\langle\langle\bm{a}^\prime|\bm{a}\rangle\rangle\langle\langle\bm{b}|}\\
        &=\sum_{\bm{b},\bm{b}^\prime\in A}{\left(\sum_{\bm{a}\in A}{{(-1)}^{\omega(\bm{a},\bm{b}+\bm{b}^\prime)}}\right)|\bm{b}^\prime\rangle\rangle\langle\langle\bm{b}|}\\
        &=|A|\sum_{\bm{b}\in A}{|\bm{b}\rangle\rangle\langle\langle\bm{b}|}=|A|\Pi_A.
    \end{split}\end{align}
The third line follows from Lemma 1 in~\citep{flammia_efficient_2020}, the fact that any Pauli commutes either with all Paulis in a subgroup of \(\mathbf{P}^n\), or exactly half of them.
Hence \(W_A\) behaves similarly to \(W\) in the sense that \(W_A\) squares to \(\Pi_A\), up to a constant, whereas \(W\) squares to the identity, the projector onto the entirety of \(\mathbf{P}^n\).

We have shown that when we attempt to estimate the eigenvalues of the gate Pauli channel \(\mathcal{E}_{ij}\) associated with the gate \(\mathcal{G}_{ij}\), we in fact estimate some of the eigenvalues of the layer Pauli channel \(\mathcal{E}_i\) associated with the layer \(\mathcal{C}_i\) in which \(\mathcal{G}_{ij}\) appears.
Taking the Walsh-Hadamard transform of these eigenvalues yields estimates for the Pauli error probabilities of \(\mathcal{E}_{ij}\) that correspond to the Pauli error probability distribution of \(\mathcal{E}_i\) marginalised onto the support of the gate \(\mathcal{G}_{ij}\).
This gives a sense in which marginalisation commutes with the Walsh-Hadamard transform.

Consequently, the ACES estimate of the noise associated with a particular gate operated in the context of a particular layer captures the averaged effect of spatial correlations within the layer on that gate.
Padding layers with single-qubit identity gates on qubits not included in the support of some existing gate in the layer ensures that we learn marginals of the full Pauli error probability distribution onto all qubits.

\section{Least squares estimation methods}\label{apdx:least-squares}

In \cref{sec:estimating-pauli-errors}, we described a weighted least squares (WLS) method for solving the standard linear regression problem of \cref{eq:least-squares}.
Alternatives include ordinary least squares (OLS) and generalised least squares (GLS).
In this appendix, we describe and compare these methods and explain our preference for WLS, drawing on standard linear regression theory described, for example, in~\citep{gelman_likelihood_2006}.

Recall \cref{eq:least-squares}, which gives the standard form linear regression problem
\begin{equation}
    \bm{b}=A\bm{x}+\bm{\epsilon},
\end{equation}
where the circuit log-eigenvalues \(\bm{b}\) are related to the gate log-eigenvalues \(\bm{x}\) by the design matrix \(A\), alongside an error variable we assume is distributed according to a multivariate normal distribution \(\bm{\epsilon}\sim\mathcal{N}_{M}{(0,\Omega^\prime)}\).
The least squares estimators for the gate log-eigenvalues in the linear regression problem \cref{eq:least-squares} can all be expressed in the form
\begin{equation}\label{eq:generalised-least-squares-estimator}
    \hat{\bm{x}}={\big(A^\intercal\hat{P}A\big)}^{-1}A^\intercal\hat{P}\bm{b},
\end{equation}
where the nature of the estimated precision matrix \(\hat{P}\) differentiates OLS, WLS, and GLS.

OLS simply sets the precision matrix to be the identity, \(\hat{P}=I\).
This corresponds to assuming that the error term \(\bm{\epsilon}\) is homoscedastic, that is, \(\epsilon_\mu\) has the same variance for each \(\mu\in M\), and uncorrelated, that is, there is zero covariance between \(\epsilon_\alpha\) and \(\epsilon_\beta\) for all \(\alpha\ne\beta\).
These assumptions entail that the covariance matrix of \(\bm{\epsilon}\) is proportional to the identity.
WLS relaxes the assumption of homoscedasticity, taking \(\hat{P}_{\mu\mu}=\hat{\Omega}_{\mu\mu}^{\prime -1}\) to be a diagonal matrix whose entries are the inverse estimates of the circuit log-eigenvalue variances.
GLS relaxes both assumptions, taking \(\hat{P}=\hat{\Omega}^{\prime -1}\) to be the inverse of the estimated circuit log-eigenvalue covariance matrix.

The accuracy of the least squares estimator improves as we progress from OLS to WLS and then to GLS, each time using a more detailed precision matrix.
Note also that GLS simplifies the expression for the gate log-eigenvalue estimator covariance matrix in \cref{eq:least-squares-covariance} to \(\Sigma^\prime={\big(A^\intercal\Omega^{\prime -1}A\big)}^{-1}\), under the assumption \(\hat{\Omega}^\prime=\Omega^\prime\) we use generally when calculating the figure of merit.

Implementing WLS and GLS requires us to estimate the diagonal elements of the circuit log-eigenvalue covariance matrix, and the full matrix \(\Omega^\prime\), respectively.
Recall that the entries of \(\Omega^\prime\) are given by \cref{eq:eigenvalue-covariance}.
We can estimate the variance of the circuit eigenvalue estimator \(\hat{\Lambda}_{T,\bm{a}}\) using that estimate for the circuit eigenvalue, making WLS easy to implement.
To construct the full covariance matrix, we may also need to calculate the off-diagonal elements \(\Lambda_{T,\bm{a}+\bm{a}^\prime}\) from the gate eigenvalues using \cref{eq:circuit-eigenvalue-design}.
Write \(A_\mu\) to denote the vector whose elements are \(A_{\mu\nu}\), the powers of the gate eigenvalues appearing in \cref{eq:circuit-eigenvalue-design} for the circuit eigenvalue \(\Lambda_\mu\).
This decomposition shows that the covariance will be zero when \(A_{T,\bm{a}}+A_{T,\bm{a}^\prime}=A_{T,\bm{a}+\bm{a}^\prime}\), and strictly positive otherwise.
The former condition is clearly satisfied when the supports of \(P_{\bm{a}}\) and \(P_{\bm{a}^\prime}\) remain disjoint throughout the action of the circuit \(\mathcal{C}_T\), removing the need to calculate \(A_{T,\bm{a}+\bm{a}^\prime}\) in that case.
It also implies that the covariance matrix is sparse even within the blocks for each tuple \(T\in\mathcal{T}\).

In this case, we cannot directly implement GLS, as the gate eigenvalues are what we are attempting to estimate.
Instead, we perform feasible generalised least squares (FGLS), first estimating the gate eigenvalues with WLS, and then iteratively generating \(\Omega^\prime\) from the estimates and performing GLS until convergence.
In practice, FGLS performs similarly to GLS with the true covariance matrix.

To efficiently numerically implement GLS, we take the sparse Cholesky factorisation \(\hat{\Omega}^\prime=LL^\intercal\) and then left-multiply the linear regression problem \cref{eq:least-squares} by \(L^{-1}\).
Indeed, as \(\Omega^\prime\) is block diagonal, we can perform this calculation separately for the blocks of \(\Omega^\prime\), which correspond to the tuples in the tuple set.

Even so, the problems of calculating the sparse Cholesky factorisation of \(\hat{\Omega}^\prime\) and inverting the Cholesky factor \(L\) are intractable for very large surface codes.
For example, at \(d=25\) there are hundreds of thousands of circuit eigenvalues.
We therefore focus on WLS in this paper, as we are primarily interested in detailing a practical and scalable protocol.
Nevertheless, GLS offers greater performance for small-scale noise characterisation experiments and should be used in these instances.
The results throughout this paper are similar for GLS.

\section{ACES estimates noise to relative precision}\label{apdx:relative-precision}

In this appendix, we optimise an experimental design for a toy circuit to demonstrate that ACES is capable of learning Pauli channels to relative precision, including in the presence of SPAM noise.
We achieve this with the usual strategy employed in tomography and noise characterisation, namely, by using a tuple that amplifies noise by repeating the layer a large number of times.

It is not generally possible to estimate all eigenvalues of a Pauli noise channel associated with a Clifford gate to relative precision~\citep{chen_learnability_2023}.
In particular, when single-qubit gates have no appreciable noise, we find that ACES is capable of relative precision estimation of the eigenvalues of those Paulis whose support is unchanged by the Clifford gate, in line with~\citep{chen_learnability_2023}.
By contrast, when single-qubit gates are appreciably noisy, we find that ACES is only capable of relative precision estimation of the eigenvalues of Paulis that commute with the Clifford gate up to sign.
This is because only the products of gate eigenvalues within each of the orbits of the gate can be learned to relative precision, as discussed in cycle error reconstruction~\citep{carignan-dugas_error_2023}.
We sidestep this issue here by considering only single-qubit Pauli gates that commute with all Paulis up to sign, though we believe that ACES is generally capable of relative precision noise estimation.

However, we caveat that this appendix demonstrates that full relative precision estimation is in fact not possible when accounting for experiment duration: it is only possible when simply considering sample complexity. 
Accounting for the time taken to implement experiments penalises the deep circuits used in relative precision noise estimation.
This results in circuits of more moderate depth---which are not capable of full relative precision noise estimation---becoming optimal.

Consider a circuit acting on \(n\) qubits that consists of a single layer of \(n\) single-qubit Pauli gates, with SPAM noise.
Each gate has \(3\) gate eigenvalues whose values we all set to \(\lambda\), and each of the \(3\) Pauli preparation and measurements on each qubit is assigned a SPAM noise eigenvalue \(\lambda_m\), so that overall we have \(N=6n\) gate eigenvalues.

We specify our tuple set as consisting of two tuples that repeat the layer \(\phi_1\) and \(\phi_2\) times, respectively.
Our experimental design is parameterised by the layer repetition numbers \(\phi_1\) and \(\phi_2\), and without loss of generality we take \(\phi_2>\phi_1\) so that we can write \(\phi_2=\phi_1+\phi\).
Note that \(\phi_1=0\) and \(\phi_2=1\) corresponds to the basic tuple set.
Ordering the circuit and gate eigenvalues appropriately yields the design matrix
\begin{equation}\label{eq:toy-design}
    A=\begin{bmatrix}
        \phi_1 I_{3n} & I_{3n} \\
        \phi_2 I_{3n} & I_{3n}
    \end{bmatrix}.
\end{equation}

Next, we need to determine the circuit eigenvalue estimator covariance matrix.
To measure all the circuit eigenvalues, each experiment merely needs to prepare and measure each of Pauli \(X\), \(Y\) and \(Z\) on the \(n\) qubits, resulting in \(|\mathcal{E}_T|=3\) experiments for any tuple \(T\).
This scheme implies that none of the circuit eigenvalue estimators are correlated with each other, producing a diagonal covariance matrix.
Now suppose that measurement and reset takes a time \(\tau\) relative to the circuit layer.
The time factor for the basic tuple set is
\begin{equation}\label{eq:toy-basic-time-factor}
    \tau_{\mathcal{T}_{\mathcal{I}}}=\frac{2\tau(\tau+1)}{2\tau+1}.
\end{equation}
Our experimental design assigns a shot weight \(\Gamma\) to the \(\phi_2\) tuple, and \(1-\Gamma\) to the \(\phi_1\) tuple, yielding a time factor
\begin{equation}\label{eq:toy-time-factor}
    \tau_{\mathcal{T},\Gamma}=\tau+\phi_1+\Gamma\phi.
\end{equation}
The measurement budget \(S\) is related to the measurement shots \(S^\prime\) as
\begin{equation}\label{eq:toy-measurement-budget}
    S=S^\prime\left(\frac{2\tau(\tau+1)}{(2\tau+1)(\tau+\phi_1+\Gamma\phi)}\right).
\end{equation}

Then, with reference to \cref{eq:eigenvalue-covariance} we see that the circuit eigenvalue estimator covariance matrix \(\Omega\) is
\begin{align}
    \Omega&=\frac{3}{S}
    \begin{bmatrix}
        \omega_1 I_{3n} & 0 \\
        0 & \omega_2 I_{3n} \\
    \end{bmatrix},\\
    \omega_1&=\frac{1-\lambda_m^2\lambda^{2\phi_1}}{1-\Gamma},\\
    \omega_2&=\frac{1-\lambda_m^2\lambda^{2\phi_2}}{\Gamma}.
\end{align}
A first-order Taylor expansion yields
\begin{align}
    \Omega^\prime&=\frac{3}{S}
    \begin{bmatrix}
        \omega_1^\prime I_{3n} & 0\\
        0 & \omega_2^\prime I_{3n}\\
    \end{bmatrix},\\
    \omega_1^\prime&=\frac{1-\lambda_m^2\lambda^{2\phi_1}}{(1-\Gamma)\lambda_m^2\lambda^{2\phi_1}},\\
    \omega_2^\prime&=\frac{1-\lambda_m^2\lambda^{2\phi_2}}{\Gamma\lambda_m^2\lambda^{2\phi_2}}.
\end{align}
As the covariance matrix is diagonal, we can treat WLS and GLS to approximate \(\hat{W}=\Omega^{\prime -1}\), simplifying \cref{eq:least-squares-covariance} to \(\Sigma^\prime={\big(A^\intercal\Omega^{\prime -1}A\big)}^{-1}\).
Calculating, and performing another first-order Taylor expansion, we obtain
\begin{align}
    \Sigma^\prime&=\frac{3}{S}
    \renewcommand{\arraystretch}{2.5}\begin{bmatrix}
        \frac{\omega_1^\prime+\omega_2^\prime}{\phi^2}I_{3n} & -\frac{\phi_2\omega_1^\prime+\phi_1\omega_2^\prime}{\phi^2}I_{3n}\\
        -\frac{\phi_2\omega_1^\prime+\phi_1\omega_2^\prime}{\phi^2}I_{3n} & \frac{\phi_2^2\omega_1^\prime+\phi_1^2\omega_2^\prime}{\phi^2}I_{3n}\\
    \end{bmatrix},\\
    \Sigma&=\frac{3}{S}
    \renewcommand{\arraystretch}{2.5}\begin{bmatrix}
        \frac{\omega_1^\prime+\omega_2^\prime}{\phi^2}\lambda^2 I_{3n} & -\frac{\phi_2\omega_1^\prime+\phi_1\omega_2^\prime}{\phi^2}\lambda_m\lambda I_{3n}\\
        -\frac{\phi_2\omega_1^\prime+\phi_1\omega_2^\prime}{\phi^2}\lambda_m\lambda I_{3n} & \frac{\phi_2^2\omega_1^\prime+\phi_1^2\omega_2^\prime}{\phi^2}\lambda_m^2 I_{3n}\\
    \end{bmatrix}.
\end{align}
The top-left block of \(\Sigma\) describes the variance of the gate eigenvalue estimators, whereas the bottom-right block describes the variance of the measurement eigenvalue estimators, and the remaining blocks describe the covariance between the gate and measurement eigenvalue estimators.

It is simplest and most instructive to focus on the large \(n\) limit.
Examining \cref{eq:aces-figure-of-merit}, and noting that both \(\tr{(\Sigma)}\) and \(\tr{(\Sigma^2)}\) are proportional to \(n\), we see that the figure of merit reduces to \(\mathcal{F}=\sqrt{S^\prime/N}\sqrt{\tr{(\Sigma)}}\).

We are interested here in relative precision noise estimation, which is only possible for the gate eigenvalues.
We will therefore consider a modified version of the figure of merit, which for convenience we still call \(\mathcal{F}\), where the trace is taken only over the top-left block of \(\Sigma\) corresponding to the gate eigenvalues.
For this modified figure of merit, we see that
\begin{align}
    \mathcal{F}&=\sqrt{\frac{(2\tau+1)(\tau+\phi_1+\Gamma\phi)}{4\tau(\tau+1)}\left(\frac{f_1}{1-\Gamma}+\frac{f_2}{\Gamma}\right)},\\
    f_1&=\frac{\lambda^2(\lambda^{-2\phi_1}-\lambda_m^2)}{\phi^2},\\
    f_2&=\frac{\lambda^2(\lambda^{-2\phi_1}\lambda^{-2\phi}-\lambda_m^2)}{\phi^2}.
\end{align}

Next, let us optimise the shot weight \(\Gamma\).
Inspecting \(\mathcal{F}\), the optimal value lies in \((0,1)\) and can be found by setting the derivative with respect to \(\Gamma\) to be zero.
This derivative is given by
\begin{equation}
    \frac{\partial\mathcal{F}}{\partial\Gamma}=\frac{2\tau+1}{8\tau(\tau+1)\mathcal{F}^\prime}\left(\frac{(\tau+\phi_2)f_1}{(1-\Gamma)^2}-\frac{(\tau+\phi_1)f_2}{\Gamma^2}\right),
\end{equation}
and is zero when
\begin{equation}
    \Gamma=\frac{\sqrt{(\tau+\phi_1)f_2}}{\sqrt{(\tau+\phi_2)f_1}+\sqrt{(\tau+\phi_1)f_2}}.
\end{equation}
Substituting and rearranging eventually yields
\begin{equation}
    \mathcal{F}=\sqrt{\frac{(2\tau+1)}{4\tau(\tau+1)}}\left(\!\sqrt{(\tau+\phi_1)f_1}+\sqrt{(\tau+\phi_2)f_2}\right).
\end{equation}

Now, we optimise the repetition number \(\phi_1\).
Differentiating with respect to \(\phi_1\) yields 
\begin{widetext}\begin{equation}
    \frac{\partial\mathcal{F}}{\partial\phi_1}=\sqrt{\frac{(2\tau+1)}{4\tau(\tau+1)}}\left(\frac{\lambda^2\big(\lambda^{-2\phi_1}\big(1-2(\tau+\phi_1)\log{(\lambda)}\big)-\lambda_m^2\big)}{2\phi^2\sqrt{(\tau+\phi_1)f_1}}+\frac{\lambda^2\big(\lambda^{-2\phi_2}\big(1-2(\tau+\phi_2)\log{(\lambda)}\big)-\lambda_m^2\big)}{2\phi^2\sqrt{(\tau+\phi_2)f_2}}\right).
\end{equation}\end{widetext}
Since \(\lambda\in(\delta,1]\) for some \(\delta>0\), we see that \(\log{(\lambda)}\) is non-positive.
Inspecting this expression, including \(f_1\) and \(f_2\), which are both always positive, it becomes clear that this derivative is positive for all \(\phi_1\).
This indicates that the optimal value of \(\phi_1\) is the minimal one, namely \(0\), as suggested by Elfving's theorem.
Then, substituting \(\phi_1\) into the expression for the figure of merit gives
\begin{equation}\begin{split}
    \mathcal{F}&=\sqrt{\frac{(2\tau+1)}{4\tau(\tau+1)}}\frac{\lambda}{\phi}\bigg(\!\sqrt{\tau(1-\lambda_m^2)}\\
    &\mspace{150mu}+\sqrt{(\tau+\phi)(\lambda^{-2\phi}-\lambda_m^2)}\bigg).
\end{split}\end{equation}

A similar calculation obtains the figure of merit when we do not account for the time taken to perform the gates, which we will call \(\mathcal{F}^\prime\), as
\begin{equation}
    \mathcal{F}^\prime=\frac{\lambda}{\sqrt{2}\phi}\left(\!\sqrt{1-\lambda_m^2}+\sqrt{\lambda^{-2\phi}-\lambda_m^2}\right).
\end{equation}
Optimising for this figure of merit \(\mathcal{F}^\prime\) produces \emph{sample-optimised} experimental designs that maximise estimation precision given a fixed number of samples from the quantum device.
By contrast, optimising for the usual figure of merit \(\mathcal{F}\) produces \emph{time-optimised} experimental designs that maximise estimation precision given a fixed window of time in which to sample from the quantum device.

Now we are ready to numerically optimise the number of repetitions \(\phi\).
To do this, we optimise it as a continuous variable and then consider both the floor and ceiling of the optimal value, choosing the one that minimises the figure of merit as the optimal repetition number \(\phi_{\mathrm{opt}}\).
For these numerical results, we choose parameter values in line with the depolarising noise model described in \cref{apdx:experimental-noise} and used elsewhere in this paper.
Namely, we choose \(\tau=660/29\) and \(\lambda_m=0.96\), and note that for single-qubit gates we usually have \(\lambda_1=0.999\), so \(1-\lambda_1=10^{-3}\).
As we focus on relative precision noise estimation, we are primarily interested in the behaviour of quantities as the gate eigenvalue \(\lambda\) approaches \(1\) from below.
We also optimise \(\phi\) for both the usual figure of merit \(\mathcal{F}\) which accounts for the time taken to perform gates, as well as \(\mathcal{F}^\prime\), which does not.

\begin{figure}[b]
    \centering
    \includegraphics[width=\columnwidth]{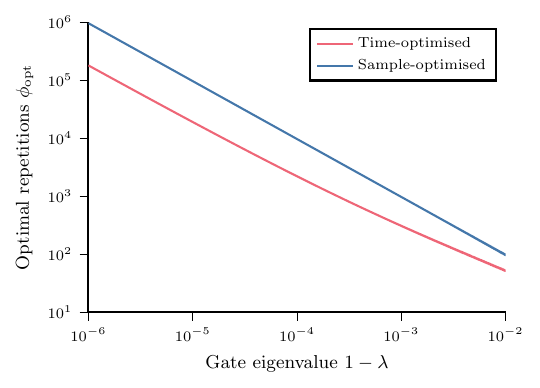}
    \vspace{-2em}
    \caption{The optimal repetition number \(\phi_{\mathrm{opt}}\) for estimating gate eigenvalues \(\lambda\) is smaller when we account for gate times than when we do not, which correspond respectively to the time- and sample-optimised experimental designs.
    Nevertheless, both optimal repetition numbers scale asymptotically as \(1/(1-\lambda)\).}
    \label{fig:toy-repetitions}
\end{figure}

\begin{figure}[t]
    \centering
    \includegraphics[width=\columnwidth]{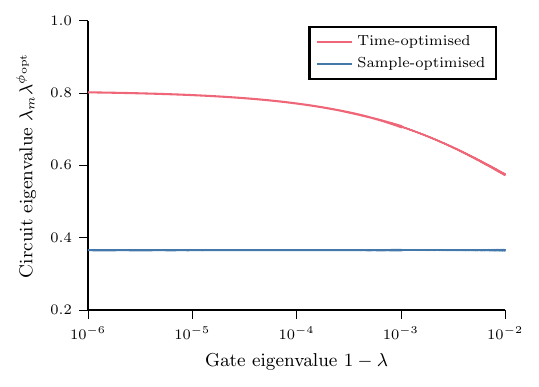}
    \vspace{-2em}
    \caption{The optimal circuit eigenvalue \(\lambda_m\lambda^{\phi_{\mathrm{opt}}}\) for estimating gate eigenvalues \(\lambda\) is larger and only slowly approaches a constant value when we account for gate times compared to when we do not.
    These correspond respectively to the time- and sample-optimised experimental designs.}
    \label{fig:toy-circuit-eigenvalue}
\end{figure}

\Cref{fig:toy-repetitions} shows the optimal repetition number \(\phi_{\mathrm{opt}}\) for estimating gate eigenvalues \(\lambda\).
Asymptotically the optimal number of repetitions scales as \(1/(1-\lambda)\), consistent with previous results in randomised benchmarking~\citep{harper_statistical_2019}, regardless of whether we account for gate times.
Accordingly, we find that the optimal circuit eigenvalue \(\lambda_m\lambda^{\phi_{\mathrm{opt}}}\) for estimating gate eigenvalues \(\lambda\) approaches a constant value, as seen in \cref{fig:toy-circuit-eigenvalue}.
The optimal circuit eigenvalue is relatively large and approached slowly when accounting for gate times, whereas when we do not, it is smaller and approximately constant regardless of \(\lambda\).

\begin{figure}[b]
    \centering
    \includegraphics[width=\columnwidth]{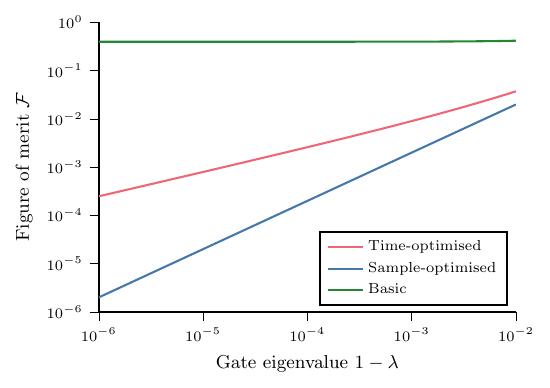}
    \vspace{-2em}
    \caption{The figure of merit \(\mathcal{F}\) of ACES noise characterisation experimental designs when estimating gate eigenvalues \(\lambda\).
    The basic experimental design is only capable of additive precision, as it does not repeat gates.
    The figures of merit of the sample- and time-optimised experimental designs scale as \(1-\lambda\) and \(\sqrt{1-\lambda}\), respectively, demonstrating that ACES is capable of relative precision noise estimation when not accounting for experiment duration.}
    \label{fig:toy-merit}
\end{figure}

Finally, \cref{fig:toy-merit} shows the figure of merit of a number of experimental designs when estimating gate eigenvalues \(\lambda\).
The figure of merit of the sample-optimised experimental design scales as \((1-\lambda)\), demonstrating that ACES is capable of relative precision noise estimation.
By contrast, the basic experimental design, which has \(\phi=1\), has a roughly constant figure of merit and is only capable of additive precision.
However, the figure of merit of the time-optimised experimental design scales as \(\sqrt{1-\lambda}\).

This toy example elucidates that full relative precision noise estimation, where estimation accuracy scales as \((1-\lambda)\), is not possible when appropriately accounting for the time taken to implement gates on the device.
Repeating gates a large number of times proportionally increases the amount of time taken to implement the circuit on a quantum device.
This decreases the number of samples collected in a fixed window of time, reducing noise estimation precision.
When accounting for this, the accuracy of the optimised experimental design instead scales as \(\sqrt{1-\lambda}\).

\section{Figure of merit differentiation}\label{apdx:merit-differentiation}

In \cref{sec:merit-gradient}, we used the derivative of the figure of merit \(\mathcal{F}\) with respect to the shot log-weights \(\gamma=\{\gamma_T\}_{T\in\mathcal{T}}\) associated with a tuple set \(\mathcal{T}\) to optimise the shot weights.
In this appendix, we detail analytic expressions for this derivative for weighted least squares as well as generalised and ordinary least squares, which are discussed in \cref{apdx:least-squares}.
This optimisation is crucial, as we find numerical automatic differentiation to be impractically slow.
We used an algorithmic tool~\citep{laue_computing_2018,laue_simple_2020} to obtain analytic expressions for the derivatives presented here, and verified them against automatic differentiation~\citep{revels_forwardmode_2016}.

Recall that the circuit eigenvalue estimator covariance matrix \(\Omega\), given by \cref{eq:eigenvalue-covariance}, is a block diagonal symmetric matrix whose blocks correspond to each tuple \(T\) in the tuple set \(\mathcal{T}\), weighted by a factor \(1/\Gamma_T\), alongside an overall experiment time factor \(\tau_{\mathcal{T},\Gamma}=\sum_{T\in\mathcal{T}}{\Gamma_T\tau_T}\).
The block diagonal structure ensures that first-order Taylor expansion used to derive the circuit log-eigenvalue estimator covariance \(\Omega^\prime\) from \(\Omega\) commutes with this weighting.
Consequently, we work with the more convenient \(\Omega^\prime\) appearing in \cref{eq:least-squares-covariance} for the gate log-eigenvalue estimator covariance matrix \(\Sigma^\prime\).
Note also that \(\Sigma=\diag{(\bm{\lambda})}\Sigma^\prime\diag{(\bm{\lambda})}\), where \(\diag{(\bm{\lambda})}\) represents the diagonal matrix created by placing the vector \(\bm{\lambda}\) along the diagonal.

First, we see that the derivative of the figure of merit with respect to the shot log-weights is related to the derivative with respect to the shot weights as
\begin{equation}
    \frac{\partial\mathcal{F}}{\partial\gamma_T}=\frac{\partial\mathcal{F}}{\partial\Gamma_U}\frac{\partial\Gamma_U}{\partial\gamma_T}=\frac{\partial\mathcal{F}}{\partial\Gamma_U}\big(\Gamma_U\Gamma_T-\delta_{UT}\Gamma_T\big).
\end{equation}
Then recalling \cref{eq:aces-figure-of-merit} for the figure of merit, applying the chain rule again yields
\begin{align}\begin{split}
        \frac{\partial\mathcal{F}}{\partial\Gamma_T}
        &=\sqrt{\frac{S^\prime}{N}}\frac{\partial\tr{(\Sigma)}}{\partial\Gamma_T}\left(\frac{\tr{(\Sigma)}^2/2+3\tr{(\Sigma^2)}/8}{\tr{(\Sigma)}^{5/2}}\right)\\
        &\quad-\sqrt{\frac{S^\prime}{N}}\frac{\partial\tr{(\Sigma^2)}}{\partial\Gamma_T}\frac{1}{4\tr{(\Sigma)}^{3/2}}.
    \end{split}
\end{align}
Hence we must determine the derivatives of \(\tr{(\Sigma)}\) and \(\tr{(\Sigma^2)}\) with respect to each of the \(\Gamma_T\).
The chain rule for matrices, as given in~\citep{petersen_matrix_2012}, yields
\begin{align}
    \frac{\partial\tr{(\Sigma)}}{\partial\Gamma_T}&=\tr{\left(\!\left(\frac{\partial\tr{(\Sigma)}}{\partial\Omega^\prime}\right)^\intercal\!\frac{\partial\Omega^\prime}{\partial\Gamma_T}\right)},\\
    \frac{\partial\tr{(\Sigma^2)}}{\partial\Gamma_T}&=\tr{\left(\!\left(\frac{\partial\tr{(\Sigma^2)}}{\partial\Omega^\prime}\right)^\intercal\!\frac{\partial\Omega^\prime}{\partial\Gamma_T}\right)}.
\end{align}
Writing the \(U\)th block of \(\Omega^\prime\) as an explicit function of the weights and measurement shots \(S^\prime\), namely \(\Omega_U^\prime=\left(\sum_{V\in\mathcal{T}}{\Gamma_{V}\tau_{V}}/\Gamma_U\right)\Omega_U^*/S^\prime\), we see that
\begin{equation}
    \frac{\partial\Omega_U^\prime}{\partial\Gamma_T}=\left(\frac{\tau_T}{\Gamma_U}-\delta_{UT}\frac{\sum_{V\in\mathcal{T}}{\Gamma_{V}\tau_{V}}}{\Gamma_T^2}\right)\frac{\Omega_U^*}{S^\prime}.
\end{equation}

Now it remains only to determine the derivatives of \(\tr{(\Sigma)}\) and \(\tr{(\Sigma^2)}\) with respect to \(\Omega^\prime\) with the help of the aforementioned algorithmic tool~\citep{laue_computing_2018,laue_simple_2020}, for ordinary, weighted, and generalised least squares estimators.

For ordinary least squares, it is helpful to define \(A_{\mathrm{OLS}}^*=\diag{(\bm{\lambda})}(A^\intercal A)^{-1}A^\intercal\), so that we have
\begin{align}
    \Sigma_{\mathrm{OLS}}
    &=A_{\mathrm{OLS}}^*\Omega^\prime A_{\mathrm{OLS}}^{*\intercal}\\
    \frac{\partial\tr{(\Sigma_{\mathrm{OLS}})}}{\partial\Omega^\prime}
    &=A_{\mathrm{OLS}}^{*\intercal}A_{\mathrm{OLS}}^*,\\
    \frac{\partial\tr{(\Sigma_{\mathrm{OLS}}^2)}}{\partial\Omega^\prime}
    &=2A_{\mathrm{OLS}}^{*\intercal}\Sigma_{\mathrm{OLS}}A_{\mathrm{OLS}}^*.
\end{align}
We obtain the same results for generalised least squares, instead defining \(A_{\mathrm{GLS}}^*=\diag{(\bm{\lambda})}(A^\intercal\Omega^{\prime -1}A)^{-1}A^\intercal\Omega^{\prime -1}\) to obtain
\begin{align}
    \Sigma_{\mathrm{GLS}}
    &=A_{\mathrm{GLS}}^*\Omega^\prime A_{\mathrm{GLS}}^{*\intercal}\\
    &=\diag{(\bm{\lambda})}(A^\intercal\Omega^{\prime -1} A)^{-1}\diag{(\bm{\lambda})},\\
    \frac{\partial\tr{(\Sigma_{\mathrm{GLS}})}}{\partial\Omega^\prime}
    &=A_{\mathrm{GLS}}^{*\intercal}A_{\mathrm{GLS}}^*,\\
    \frac{\partial\tr{(\Sigma_{\mathrm{GLS}}^2)}}{\partial\Omega^\prime}
    &=2A_{\mathrm{GLS}}^{*\intercal}\Sigma_{\mathrm{GLS}}A_{\mathrm{GLS}}^*.
\end{align}
However, weighted least squares is a little more complicated.
We define the diagonal weight matrix with the elementwise Hadamard product as \(W=(\Omega^\prime\odot I)^{-1}\), and then both \(A_{\mathrm{WLS}}^+=(A^\intercal WA)^{-1}A^\intercal W\) and \(A_{\mathrm{WLS}}^*=\diag{(\bm{\lambda})}A_{\mathrm{WLS}}^+\), before finally defining \(B_{\mathrm{WLS}}=\Omega^\prime W(AA_{\mathrm{WLS}}^+ - I)\).
Together, these yield
\begin{align}
    \Sigma_{\mathrm{WLS}}
    &=A_{\mathrm{WLS}}^*\Omega^\prime A_{\mathrm{WLS}}^{*\intercal},\\
    \frac{\partial\tr{(\Sigma_{\mathrm{WLS}})}}{\partial\Omega^\prime}
    &=A_{\mathrm{WLS}}^{*\intercal}A_{\mathrm{WLS}}^*\\
    &\quad+2A_{\mathrm{WLS}}^{*\intercal}A_{\mathrm{WLS}}^*B_{\mathrm{WLS}}\odot I,\\
    \frac{\partial\tr{(\Sigma_{\mathrm{WLS}}^2)}}{\partial\Omega^\prime}
    &=2A_{\mathrm{WLS}}^{*\intercal}\Sigma_{\mathrm{WLS}}A_{\mathrm{WLS}}^*\\
    &\quad+4A_{\mathrm{WLS}}^{*\intercal}\Sigma_{\mathrm{WLS}}A_{\mathrm{WLS}}^*B_{\mathrm{WLS}}\odot I.
\end{align}
Note that the WLS estimator \(A_{\mathrm{WLS}}^+\) is a left inverse to \(A\), such that \(AA_{\mathrm{WLS}}^+-I=[A,A_{\mathrm{WLS}}^+]\) is in fact a commutator.
This commutator is zero when \(A\) is square and full-rank and therefore admits a right inverse, as in the case of the basic experimental design.

\section{Tuple generation}\label{apdx:tuple-generation}

When generating optimised tuple sets in \cref{sec:tuple-optimisation}, there are two stages at which we must generate tuples.
First, we generate a set \(\mathcal{T}_{\mathrm{rep}}\) of tuples which will be repeated some large number of times to amplify and estimate small gate eigenvalues.
These resemble the germs of gate set tomography~\citep{nielsen_gate_2021}.
We later append random tuples to the tuple set, drawing them from some probability distribution over tuples \(\mathcal{P}\).
In this appendix, we describe how we generate both these repeated and random tuples.
We expect it is possible to substantially improve upon the heuristic methods outlined here.

Our tuple generation methods depend on whether the circuit \(\mathcal{C}\) is dynamically decoupled.
We would like both the repeated and random tuples to reflect this property so as to more closely reflect the original circuit.
We find it performant to repeat the layers comprising the repeated tuples such that each repeated tuple implements an involution, and repeating it twice implements the identity.
The repeated tuples for non-dynamically-decoupled circuits are simple and consist of each of the non-empty tuples of the basic tuple set \(\mathcal{T}_{\mathcal{I}}\).
For dynamically decoupled circuits, single-qubit gate layers are treated in the same manner, whereas layers with multi-qubit gates are interleaved with the dynamical decoupling layer.
This ensures that multi-qubit gate layers are not performed in succession, as is the case in the surface code syndrome extraction circuit we aim to characterise.
For a concrete example, see the tuples in \cref{tab:optimised-design}.

When generating random tuples, which we do by iteratively appending random elements to the tuple, we will often use a discrete power-law distribution with finite support known as the \emph{generalised Zipf distribution}.
Up to a normalisation factor, this distribution assigns a weight \(1/u^s\) to an outcome \(u\) up to a maximum value \(u_{\mathrm{max}}\).
We will also consider mirror circuits, which implement a series of layers followed by their inverses in reverse order, such that the overall circuit implements the identity.
These have recently been used to improve the efficiency of randomised benchmarking~\citep{mayer_theory_2023, proctor_scalable_2022}.
As surface code syndrome extraction circuits consist of layers of \(X\), Hadamard and controlled-\(X\) or controlled-\(Z\) gates, which are all their own inverses, mirror circuits are generated by mirror tuples, with a simple example of a mirror tuple being \(T=(2,1,3,2,2,3,1,2)\).
In the context of ACES, we must append additional layers to mirror circuits in order to obtain full-rank design matrices~\citep{flammia_averaged_2022}, and accordingly we mirror only the first \(L^\prime=\lfloor (L-1)/2\rfloor\) elements of a length \(L\) tuple, leaving one or two non-mirror layers at the end of the tuple for odd or even \(L\), respectively.

To generate a random tuple, we first choose the tuple length \(L\) randomly according to a Zipf distribution with \(s=1\) and a maximum length \(u_{\mathrm{max}}=2l\) of twice the length \(l\) of the circuit \(\mathcal{C}\).
We randomly choose whether the tuple will be mirrored or not, and the rule for the number of copies of each index we append to the tuple, each with even probability.
Specifically, we append a Zipf-distributed number of copies where we choose either \(s=\infty\), which corresponds to appending a single copy, or \(s=2\), which has some probability of appending multiple copies.

For non-dynamically-decoupled circuits, we then simply iteratively append copies of a random index drawn with even probability from the unique layer indices \(\mathcal{I}\).
It is a little more complicated for dynamically decoupled circuits, as we again want to avoid repeating multi-qubit gate layers.
We do this by appending copies of pairs of indices, where the first is drawn before the second to ensure that multi-qubit gate layers are indeed not repeated.
When a pair is repeated, the Zipf-distributed repetition number is divided by \(2\) and rounded up.
In the case of mirror circuits, we must also check that a multi-qubit gate layer is not repeated at the point of mirroring.

This heuristic range of rules for generating tuples ensures that we generate a wide variety of random tuples.
Despite this, we find that our optimisation algorithm \cref{alg:optimise-tuple-set} generally chooses short random tuples.
This is driven at least in part by the fact that the number of repetitions for \(\mathcal{T}_{\mathrm{rep}}\) is optimised when augmented by the basic tuple set \(\mathcal{T}_{\mathrm{I}}\).
We also suspect that the efficient packing of circuit eigenvalue measurements, as specified by \cref{alg:pack-experiment-set}, is a key driver of performance, and this is achieved by mirror tuples, tuples with repeated layers, and shallow tuples.
A deeper theoretical understanding of tuple sets should suggest improvements to these methods.

\section{An experimentally-relevant noise model}\label{apdx:experimental-noise}

We would like to examine the performance of our methods on an experimentally-relevant noise model.
In this appendix, we construct a distribution over random noise models which resemble and are qualitatively similar to the noise observed in the surface code experiment~\citep{acharya_suppressing_2023}, whose syndrome extraction circuit we depicted in \cref{fig:rotated-code-circuit}.
We call this distribution log-normal Pauli noise because each non-identity Pauli and measurement error probability is randomly and independently distributed according to some log-normal distribution.

First, we specify the parameters we use for the gate layer, measurement, and reset times.
In~\citep{acharya_suppressing_2023}, the overall syndrome extraction circuit is specified to take \(921\) ns, with measurement taking \(500\) ns and reset taking \(160\) ns, leaving \(261\) ns for the rest of the circuit.
Gate layer times are not obviously specified, so we assume they are equal, which is not entirely inconsistent with an earlier experiment~\citep{chen_exponential_2021}.
This yields a single-qubit gate layer time \(\tau_1=29\textrm{ ns}\), a two-qubit gate layer time \(\tau_2=29\textrm{ ns}\), and a measurement and reset time \(\tau_{\mathrm{mr}}=660\textrm{ ns}\).

For each gate type, including measurements, we tune the mean and variance of the log-normal distribution such that the entanglement infidelity---the sum of a gate's non-identity Pauli error probabilities---has a mean and variance resembling~\citep{acharya_suppressing_2023}.
In particular, we target mean gate infidelities of \(r_1=0.075\%\) for single-qubit gates, \(r_2=0.5\%\) for two-qubit gates, and \(r_m=2\%\) for measurements.
Note that the single-qubit identity gates used to pad layers are treated the same as other single-qubit gates.
Approximating the sum of log-normal distributions as another log-normal distribution with the same mean and variance, we arbitrarily specify that the normal random variable underlying this new distribution has a not dissimilar variance \(\sigma_{\mathrm{tot}}^2=\log{(10/9)}\approx 0.105\).

A log-normally distributed random variable \(X=\exp{(\mu_Z+\sigma_Z Z)}\), where \(Z\) is a normal random variable with mean \(0\) and variance \(1\), such that \(\mu_Z+\sigma_Z Z\) has mean \(\mu_Z\) and variance \(\sigma_Z^2\).
Then it is well-known that \(X\) has mean \(\mu_X=\exp{(\mu_Z+\sigma_Z^2/2)}\) and variance \(\sigma_X^2=\big(\exp{(\sigma_Z^2)}-1\big)\exp{(2\mu_Z+\sigma_Z^2)}\).
We begin by specifying the distribution used to generate measurement error probabilities, as each measurement only has one error probability.
Setting \(\mu_X=r_m\) and \(\sigma_Z^2=\sigma_{\mathrm{tot}}^2\), rearranging yields \(\mu_Z=\log{(\mu_X)}-\sigma_Z^2/2=-\log{(50)}-\log{(10/9)}/2\approx -3.965\).

Gates are a little more complicated, as a \(b\)-qubit gate has \(b^\prime=4^b-1\) non-identity Pauli error probabilities.
We therefore approximate the sum of these \(b^\prime\) log-normal distributions as log-normally distributed with the same mean \(\mu_{X;b}=b^\prime \mu_X\) and variance \(\sigma_{X;b}^2=b^\prime \sigma_X^2\).
Simple rearrangements yield the mean \(\mu_Z\) and variance \(\sigma_Z^2\) of the normal random variable underlying each individual non-identity Pauli error probability as
\begin{align}
    \sigma_Z^2&=\log{\big(1+b^\prime(\exp{(\sigma_{Z;b}^2)}-1)\big)},\\
    \mu_Z&=\log{\big(\mu_{X;b}/b^\prime\big)}-\sigma_Z^2/2.
\end{align}

For single-qubit gates where \(b=1\), we set \(\sigma_{Z;1}^2=\sigma_{\mathrm{tot}}^2\) and \(\mu_{X;1}=r_1\), yielding \(\sigma_Z^2=\log{(4/3)}\approx 0.288\) and \(\mu_Z=-\log{(8000/\sqrt{3})}\approx -8.438\).
Similarly for two-qubit gates where \(b=2\), we set \(\mu_{X;2}=r_2\) and \(\sigma_{Z;2}^2=\sigma_{\mathrm{tot}}^2\), yielding \(\sigma_Z^2=\log{(8/3)}\approx 0.981\) and \(\mu_Z=-\log{(2000\sqrt{6})}\approx -8.497\).
As the sum of log-normal distributions is not in fact log-normal, the true variance of the sum will be a little less than our target, with greater deviation for two-qubit gates than for single-qubit gates.
Nevertheless, the mean will be consistent with our target.

We can easily compare this to a depolarising Pauli noise model with the same gate and measurement entanglement infidelities \(r_2\), \(r_1\), and \(r_m\).
We could think of this as being produced by setting \(\sigma_{\mathrm{tot}}^2=0\).
Be careful to note that under the usual parameterisation, a \(b\)-qubit depolarising channel with entanglement infidelity \(r_b\) in fact has a depolarising constant \(r_b(1+1/(4^b-1))\).

It is easier to calculate quantities for depolarising noise, as they can be calculated directly rather than estimated across random samples of log-normal Pauli noise.
Nevertheless, random instances of log-normal Pauli noise are a more interesting and representative target for noise characterisation.

\newpage

\begin{table}[H]
    \centering
    \vspace{-1.25em}
    \caption{The experimental design with \(31\) tuples optimised for depolarising noise for the syndrome extraction circuit of a distance-\(3\) surface code shown in \cref{fig:rotated-code-circuit}, including the measurement shot weights \(\Gamma\) and tuple set \(\mathcal{T}\), as well as the repetition number for the repeated tuples.}
    \vspace{0.25em}
    \def\arraystretch{1.4}
    \footnotesize
    \begin{tabular}{|c|c|c}
        \hline
        Shot weight & Tuple        & \multicolumn{1}{c|}{Repetition number} \\
        \hline
        0.000516    & (1)          & \multicolumn{1}{c|}{233} \\
        0.000506    & (3)          & \multicolumn{1}{c|}{233} \\
        0.002424    & (5)          & \multicolumn{1}{c|}{191} \\
        0.007184    & (2, 5, 2, 5) & \multicolumn{1}{c|}{25}  \\
        0.007167    & (4, 5, 4, 5) & \multicolumn{1}{c|}{25}  \\
        0.007138    & (6, 5, 6, 5) & \multicolumn{1}{c|}{25}  \\
        0.007317    & (8, 5, 8, 5) & \multicolumn{1}{c|}{25}  \\
        \hline
        0.158506    & (2)          & \\
        0.204358    & (4)          & \\
        0.139123    & (6)          & \\
        0.074416    & (8)          & \\
        0.030260    & (1, 4)       & \\
        0.022677    & (1, 6)       & \\
        0.029468    & (1, 8)       & \\
        0.058915    & (2, 1)       & \\
        0.002187    & (2, 5)       & \\
        0.045958    & (3, 6)       & \\
        0.029315    & (3, 8)       & \\
        0.003016    & (5, 2)       & \\
        0.002618    & (5, 4)       & \\
        0.002297    & (5, 8)       & \\
        0.035342    & (6, 1)       & \\
        0.032844    & (8, 1)       & \\
        0.078418    & (8, 3)       & \\
        0.011389    & (2, 5, 3)    & \\
        0.003830    & (3, 2, 5)    & \\
        0.001268    & (4, 1, 5)    & \\
        0.000428    & (4, 5, 5)    & \\
        0.000323    & (5, 5, 4, 1) & \\
        0.000505    & (5, 5, 6, 3) & \\
        0.000466    & (5, 6, 3, 5) & \\
        \cline{1-2}
    \end{tabular}
    \label{tab:optimised-design}
\end{table}

\newpage

\section{Additional numerical results}\label{apdx:add-results}

We present a number of additional numerical results that were not included in \cref{sec:numerical-results}.
To further examine these results, refer to the code~\citep{hockings_quantumacesjl_2025}.

\begin{figure}[b]
    \centering
    \includegraphics[width=\columnwidth]{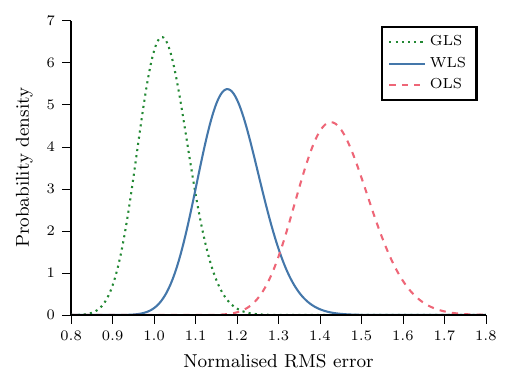}
    \vspace{-2em}
    \caption{Performance comparison of experimental designs optimised for generalised least squares (GLS), weighted least squares (WLS), and ordinary least squares (OLS).
    Predicted normalised RMS error probability distributions for a fixed-seed random instance of log-normal Pauli noise for the syndrome extraction circuit of a distance-\(3\) surface code.}
    \label{fig:ls-nrmse-pdfs}
\end{figure}

The experimental design optimised for depolarising noise on the syndrome extraction circuit of a distance-\(3\) surface code, which we focus on in \cref{sec:numerical-results}, is displayed in \cref{tab:optimised-design}.
Recall that the elements of the tuples index the gate layers appearing in \cref{fig:rotated-code-circuit}.
We see that while the repeated tuples are very long, the other tuples are very shallow, as suggested by Elfving's theorem.
We can also compare our repetitions to the toy example of \cref{apdx:relative-precision} to check the performance of the repetition number optimisation algorithm.
There, the optimal number of Pauli gate repetitions at the same parameter values is 306, compared to 191 here, although only 176 repetitions are required to achieve the asymptotically optimal circuit eigenvalue.
These values are similar enough so as to not be a cause for concern, given the different contexts.

\begin{figure*}[t!]
    \centering
    \includegraphics[width=\textwidth]{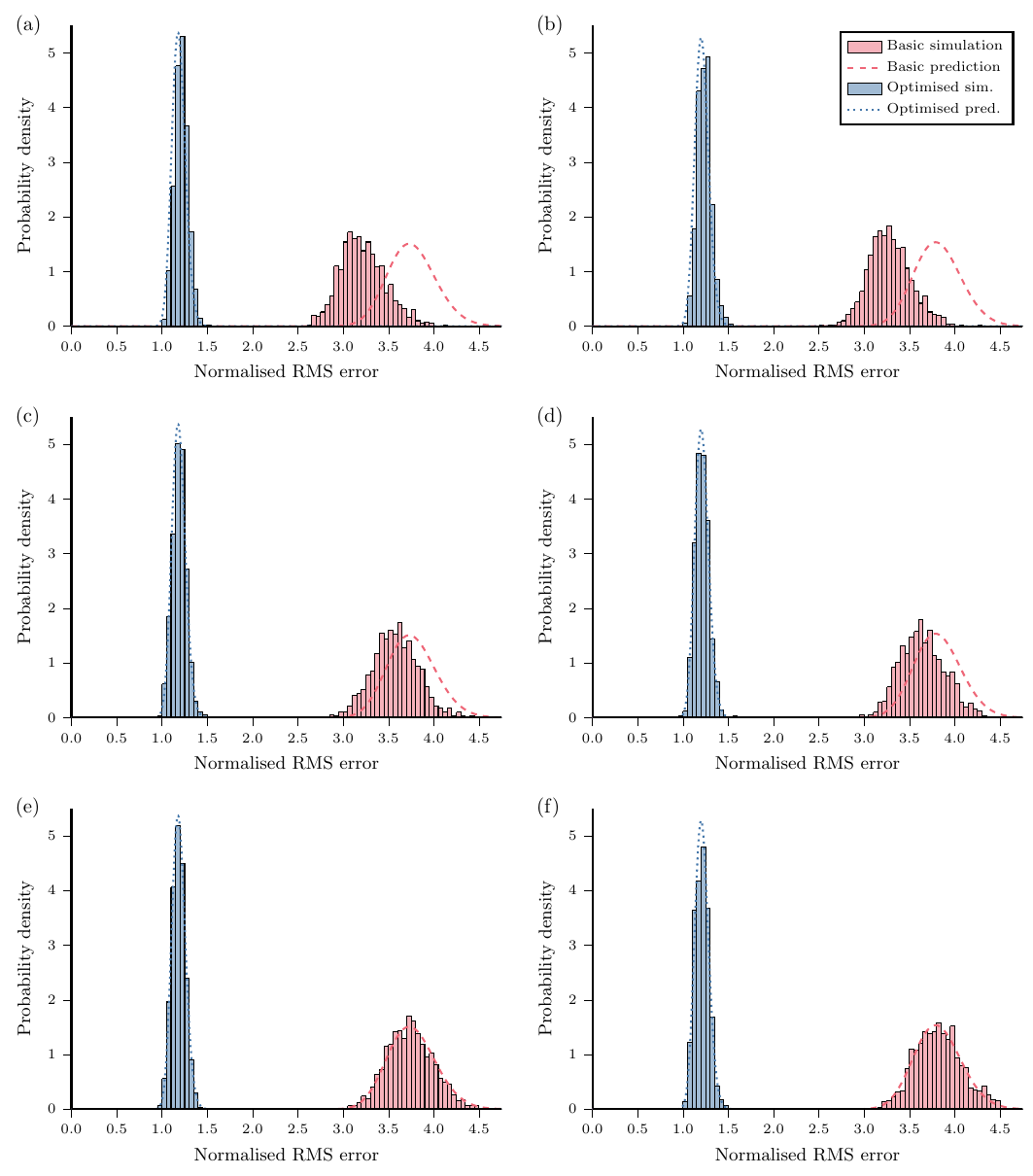}
    \vspace{-2em}
    \caption{Similar results as in \cref{fig:aces-performance} (reproduced in (e)), but across a range of measurement budgets \(S\) and for both a fixed-seed random instance of log-normal Pauli noise and depolarising noise.
    Specifically, (a, c, e), log-normal Pauli noise; (b, d, f) depolarising noise; (a, b) \(S=10^6\); (c, d) \(S=10^7\); and (e, f) \(S=10^8\).}
    \label{fig:aces-performance-hists}
\end{figure*}

\begin{figure*}[t!]
    \centering
    \includegraphics[width=\textwidth]{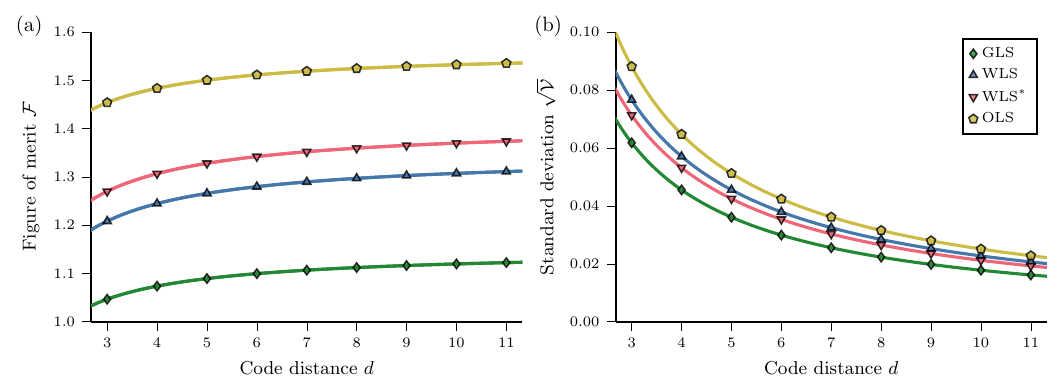}
    \vspace{-2em}
    \caption{Similar results as in \cref{fig:merit-distance-scaling}, only for depolarising noise but across a range of experimental designs.
    The experimental designs are optimised for generalised least squares (GLS), weighted least squares (WLS), and ordinary least squares (OLS), and include one optimised for inappropriate average error rates (WLS*), namely the worst performing design in \cref{fig:merit-heatmap}.
    (a) Figure of merit \(\mathcal{F}\), the expected normalised RMS error, and (b) the normalised RMS error standard deviation \(\sqrt{\mathcal{V}}\) as functions of the surface code distance \(d\), with the data points fit by the same functional forms as in \cref{fig:merit-distance-scaling}.}
    \label{fig:merit-distance-scaling-combined}
\end{figure*}

Interestingly, most of the measurement budget goes towards measuring the controlled-\(Z\) gate layers by themselves, and this is preferred over measuring them alongside single-qubit gate layers.
Moreover, no tuple measures multiple controlled-\(Z\) gate layers together.
This illustrates that performant noise characterisation experimental designs are in fact quite simple.

\Cref{fig:ls-nrmse-pdfs} compares the predicted performance for experimental designs optimised for the different least squares estimators.
As we would expect, GLS performs best, followed by WLS, and then OLS performs worst.
Indeed, the expected figures of merit across \(400\) random instances of log-normal Pauli noise are \(1.0346\pm 0.0011\), \(1.2001\pm 0.0014\), and \(1.4601\pm 0.0017\), respectively.
These performance differences are much more substantial than for the designs optimised for inappropriate average error rates.

\begin{table}[b]
    \centering
    \vspace{-1.25em}
    \caption{Condition numbers of the design matrices of optimised experimental designs at a range of surface code distances.
    The experimental designs are optimised for generalised least squares (GLS), weighted least squares (WLS), and ordinary least squares (OLS), and include the basic design as well as one optimised for inappropriate average error rates (WLS*), namely the worst performing design in \cref{fig:merit-heatmap}.}
    \vspace{0.25em}
    \def\arraystretch{1.4}
    \begin{tabular}{|c|c|c|c|c|c|}
        \hline
        Code distance & GLS & WLS & WLS* & OLS & Basic \\
        \hline
        3  & 187.98 & 181.75 & 380.56 & 97.57 & 29.39 \\
        4  & 191.85 & 185.86 & 387.92 & 98.01 & 30.97 \\
        5  & 193.86 & 188.13 & 391.98 & 98.23 & 31.70 \\
        6  & 195.17 & 189.59 & 394.67 & 98.35 & 32.09 \\
        7  & 196.13 & 190.63 & 396.61 & 98.43 & 32.33 \\
        8  & 196.85 & 191.39 & 398.07 & 98.49 & 32.48 \\
        9  & 197.40 & 191.96 & 399.19 & 98.53 & 32.59 \\
        10 & 197.83 & 192.41 & 400.07 & 98.55 & 32.66 \\
        11 & 198.18 & 192.77 & 400.77 & 98.57 & 32.71 \\
        \hline
    \end{tabular}
    \label{tab:design-cond-num}
\end{table}

\begin{table}[b]
    \centering
    \vspace{-1.25em}
    \caption{Pseudoinverse norms of the design matrices of optimised experimental designs at a range of surface code distances.
    The experimental designs are optimised for generalised least squares (GLS), weighted least squares (WLS), and ordinary least squares (OLS), and include the basic design as well as one optimised for inappropriate average error rates (WLS*), namely the worst performing design in \cref{fig:merit-heatmap}.}
    \vspace{0.25em}
    \def\arraystretch{1.4}
    \begin{tabular}{|c|c|c|c|c|c|}
        \hline
        Code distance & GLS & WLS & WLS* & OLS & Basic \\
        \hline
        3  & 0.5994 & 0.5838 & 0.5435 & 0.7061 & 5.4211 \\
        4  & 0.6002 & 0.5856 & 0.5446 & 0.7064 & 5.5647 \\
        5  & 0.6013 & 0.5876 & 0.5460 & 0.7066 & 5.6301 \\
        6  & 0.6026 & 0.5894 & 0.5475 & 0.7068 & 5.6651 \\
        7  & 0.6039 & 0.5910 & 0.5488 & 0.7069 & 5.6859 \\
        8  & 0.6050 & 0.5923 & 0.5499 & 0.7070 & 5.6993 \\
        9  & 0.6060 & 0.5933 & 0.5509 & 0.7071 & 5.7085 \\
        10 & 0.6068 & 0.5942 & 0.5517 & 0.7072 & 5.7149 \\
        11 & 0.6075 & 0.5949 & 0.5523 & 0.7072 & 5.7197 \\
        \hline
    \end{tabular}
    \label{tab:design-pinv-norm}
\end{table}

Then, we perform ACES noise characterisation on the same fixed-seed random instance of log-normal Pauli noise as in \cref{fig:aces-performance}, as well as depolarising noise, comparing the basic and optimised experimental designs.
\Cref{fig:aces-performance-hists} show the distribution of the normalised RMS error over \(1000\) simulated trials of ACES across measurement budgets \(S\in\{10^6,10^7,10^8\}\) for both noise models.
These figures also show the predicted normalised RMS error distributions, which closely align with the simulated data for the largest measurement budget \(S=10^8\) but are overly pessimistic for smaller measurement budgets.
As discussed previously, this is because we ensure the parameter estimates are within bounds, which improves performance at small measurement budgets.

Next, we examine how the performance of these optimised experimental designs for depolarising noise varies as a function of the surface code distance \(d\).
Specifically, \cref{fig:merit-distance-scaling-combined} shows the figure of merit, the expected normalised RMS error, and the normalised RMS error standard deviation, for the GLS-, WLS-, and OLS-optimised experimental designs and their respective estimators. 
It also includes the worst performing WLS-optimised experimental design from \cref{fig:merit-heatmap}.
The functional forms again fit the data well, although the GLS figure of merit has much larger relative errors which nevertheless remain under \(10^{-3}\) for each data point.
Perhaps this is because the experiment number for the GLS-optimised experimental design is not constant across all distances \(d\), unlike the other designs.
Notably, the relative errors for the basic design, which is not shown here, remain under \(10^{-14}\) for each data point and approach numerical precision.

In \cref{tab:design-cond-num,tab:design-pinv-norm}, we show the condition numbers and pseudoinverse norms of the design matrices of these experimental designs at a range of surface code distances \(d\).
As with the figure of merit, the condition number and pseudoinverse norm do not substantially increase with the code distance \(d\).
Note that at each distance, the condition number for the basic design is the square of the pseudoinverse norm.
The pseudoinverse norm was proposed as a candidate figure of merit in~\citep{flammia_averaged_2022}.
Comparing the optimised and basic experimental designs, we see that the optimisation algorithm decreases the pseudoinverse norm at the cost of increasing the condition number, suggesting that the pseudoinverse norm is a superior figure of merit.
However, our figure of merit precisely describes the performance of experimental designs and does not always correlate with the pseudoinverse norm.
For example, we see in \cref{tab:design-pinv-norm} that the worst performing design in \cref{fig:merit-heatmap} has a better pseudoinverse norm than the best performing design.

\begin{figure}[t]
    \centering
    \includegraphics[width=\columnwidth]{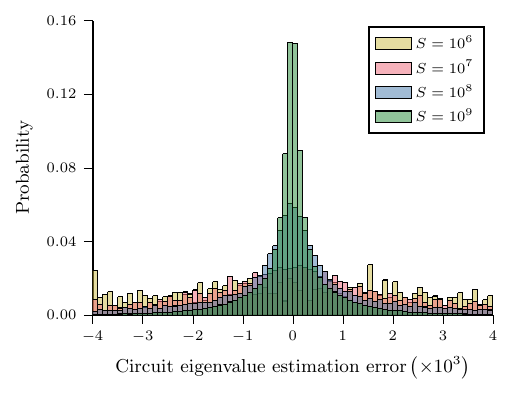}
    \vspace{-2em}
    \caption{Similar results as in \cref{fig:aces-eigenvalue-histogram}, but characterising depolarising noise rather than a fixed-seed random instance of log-normal Pauli noise.
    The circuit eigenvalue estimation error distributions are shown for the distance-\(25\) surface code syndrome extraction circuit.}
    \label{fig:aces-eigenvalue-histogram-depolarising}
\end{figure}

\begin{figure*}[t]
    \centering
    \includegraphics[width=\textwidth]{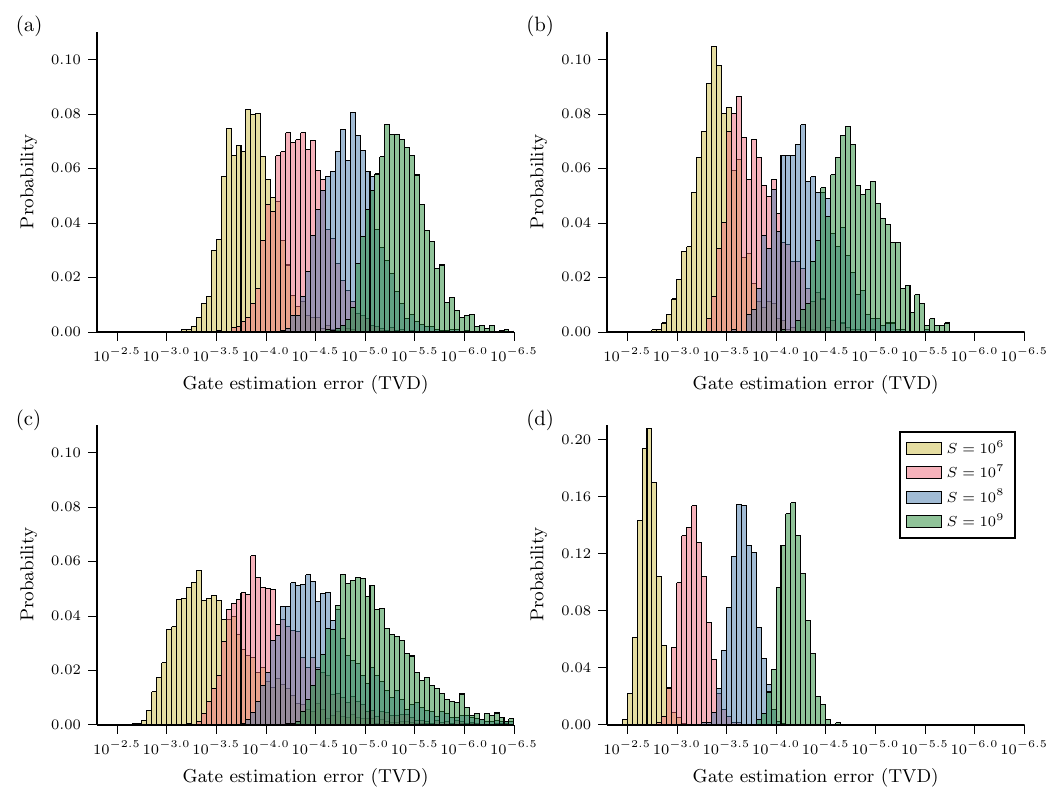}
    \vspace{-2em}
    \caption{Similar results as in \cref{fig:aces-tvd-histograms}, but characterising depolarising noise rather than a fixed-seed random instance of log-normal Pauli noise.
    The distributions of the gate estimation error are shown across the range of gate types appearing in the distance-\(25\) surface code syndrome extraction circuit, namely: (a) dynamical decoupling \(X\) gates and padded identity gates, or Pauli gates; (b) Hadamard gates; (c) measurements; and (d) controlled-\(Z\) gates.
    Results are shown for the optimised experimental design over a range of measurement budgets \(S\).}
    \label{fig:aces-tvd-histograms-depolarising}
\end{figure*}

Finally, we show the optimised experimental design characterising depolarising noise on the syndrome extraction circuit of a distance-\(25\) surface code.
The results are similar to those for a fixed-seed random instance of log-normal Pauli noise.
Here, the fits in \cref{fig:merit-distance-scaling} predict a figure of merit, or expected normalised RMS error, of \(1.3318\), with a normalised RMS error standard deviation of \(0.0091\).
In comparison, the normalised RMS errors were \(\{1.3716,1.3349,1.3215,1.3196\}\) across the measurement budgets \(S\in\{10^6,10^7,10^8,10^9\}\), which are consistent with the prediction excepting, again, the smallest measurement budget.
\Cref{fig:aces-eigenvalue-histogram-depolarising} shows the distribution of the circuit eigenvalue estimation error, and \cref{fig:aces-tvd-histograms-depolarising} shows the distribution of the gate estimation error, expressed as the total variation distance (TVD), across the range of gate types appearing in the circuit.
We again observe the expected scaling of the median gate estimation errors with the measurement budget \(S\), quantified here in \cref{tab:median-tvd-depolarising}, with improved performance at small measurement budgets as we ensure parameter estimates are within bounds.

\begin{table}[t]
    \centering
    \vspace{-1.25em}
    \caption{Similar results as in \cref{tab:median-tvd}, but instead showing the medians of the gate estimation error distributions in \cref{fig:aces-tvd-histograms-depolarising}.}
    \vspace{0.25em}
    \def\arraystretch{1.4}
    \begin{tabular}{|c|c|c|c|c|}
        \hline
        \multirow[c]{2}{5.75em}{\centering Measurement budget} & \multicolumn{4}{c|}{\(-\log_{10}\) median gate estimation error (TVD)} \\
        \cline{2-5}
                 & Pauli & Hadamard & Measurement & Controlled-\(Z\) \\
        \hline
        \(10^6\) & 3.847 & 3.409    & 3.479       & 2.719            \\
        \(10^7\) & 4.345 & 3.782    & 4.029       & 3.164            \\
        \(10^8\) & 4.850 & 4.285    & 4.541       & 3.666            \\
        \(10^9\) & 5.336 & 4.790    & 5.045       & 4.168            \\
        \hline
    \end{tabular}
    \label{tab:median-tvd-depolarising}
\end{table}

\clearpage

\section{ACES for unrotated surface codes}\label{apdx:unrotated-surface-codes}

Our numerical results have focused on the syndrome extraction circuit of rotated surface codes.
However, we believe our methods apply to the syndrome extraction circuits of arbitrary topological quantum codes.
To this end, here we examine the syndrome extraction circuit of unrotated surface codes.
Our circuit consists of four layers of controlled-\(X\) gates sandwiched between two identical layers of Hadamard gates, such that there are six layers in the circuit and five unique circuit layers.
The circuit follows the circuit given in~\citep{tomita_lowdistance_2014} and does not feature dynamical decoupling.

In this appendix, we demonstrate that our methods are also able to characterise the syndrome extraction circuits of unrotated surface codes by reproducing the results in \cref{sec:numerical-results}.
We obtain remarkably similar results to those for the rotated surface code, despite not modifying our method or parameters.
The additional results presented in \cref{apdx:add-results} are also similar for the unrotated surface code, and although we do not reproduce them here, they can be found with the code~\citep{hockings_quantumacesjl_2025}.

\begin{figure}[b]
    \centering
    \includegraphics[width=\columnwidth]{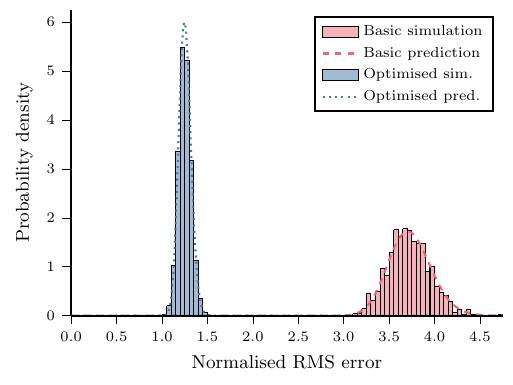}
    \vspace{-2em}
    \caption{Similar results as \cref{fig:aces-performance}, but for the unrotated surface code.
    Simulated data show the distribution of the normalised RMS error between the estimated and true gate eigenvalues when characterising a fixed-seed random instance of log-normal Pauli noise for the syndrome extraction circuit of a distance-\(3\) unrotated surface code.
    The data are \(1000\) trials of characterising a fixed-seed random instance of log-normal Pauli noise with a measurement budget \(S=10^8\) and align with the predicted performance distributions.}
    \label{fig:unrotated-aces-performance}
\end{figure}

\begin{figure*}[t]
    \centering
    \includegraphics[width=\textwidth]{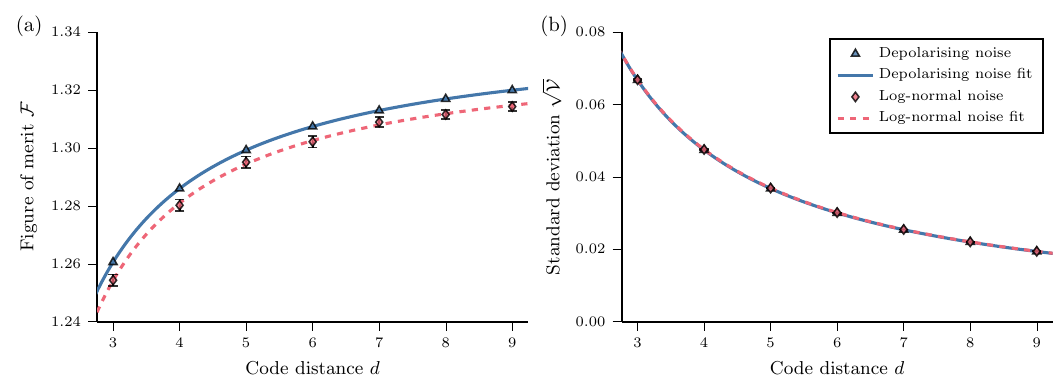}
    \vspace{-2em}
    \caption{Similar results as in \cref{fig:merit-distance-scaling}, but for the unrotated surface code.
    (a) Figure of merit \(\mathcal{F}\), the expected normalised RMS error, and (b) the normalised RMS error standard deviation \(\sqrt{\mathcal{V}}\), of the optimised experimental design for the syndrome extraction circuit, shown as functions of the unrotated surface code distance \(d\).
    The data points have no error for depolarising noise, whereas for log-normal Pauli noise, they report the mean estimated by sampling from the distribution over noise models, with error bars indicating two standard deviations.
    The functional forms fit the data precisely, with relative errors of under \(10^{-9}\) for each depolarising noise data point.}
    \label{fig:unrotated-merit-distance-scaling}
\end{figure*}

We first optimise an experimental design for depolarising noise on the syndrome extraction circuit of a distance-\(3\) unrotated surface code on \(25\) qubits, using the procedures outlined in \cref{sec:designing-aces}.
The random tuples added by \cref{alg:optimise-tuple-set} are shallow, with depth at most \(4\), and the shot weights allocate the majority of the measurement budget to individually measuring the controlled-\(X\) gate layers.

Then we perform ACES noise characterisation on a fixed-seed random instance of log-normal Pauli noise, comparing the basic and optimised experimental designs.
\Cref{fig:unrotated-aces-performance} shows the distribution of the normalised RMS error over \(1000\) simulated trials of ACES with a measurement budget \(S=10^8\) for both designs, as well as the predicted normalised RMS error distributions.
The predicted distributions closely align with the simulated data, validating our ability to predict the performance of ACES noise characterisation experiments, and demonstrating that the optimised design substantially outperforms the basic design.
For this random instance of log-normal Pauli noise, the basic design has a figure of merit, or expected normalised RMS error, that is larger than the optimised design by a factor \(2.98\), implying that the sample efficiency of the optimised design is a factor \(8.86\) better than the basic design.

Next, we examine how the performance of this optimised experimental design for the syndrome extraction circuit varies as a function of the unrotated surface code distance \(d\).
\Cref{fig:unrotated-merit-distance-scaling} depicts the figure of merit \(\mathcal{F}\), the expected normalised RMS error, and the normalised RMS error standard deviation \(\sqrt{\mathcal{V}}\) as functions of the unrotated surface code distance \(d\).
The exact quantities are reported for depolarising noise, whereas for log-normal Pauli noise, we report estimates of the mean across random samples from the distribution over noise models.

Empirically, we find that for depolarising noise, the normalised trace of the gate eigenvalue estimator covariance matrix \(\tr{(\Sigma)}/S^\prime\), and of its square \(\tr{(\Sigma^2)}/S^{\prime 2}\), are precisely fit as quadratic functions of \(d\).
Model selection using the Akaike information criterion corrected for small samples~\citep{hurvich_regression_1989}, prefers this quadratic model over other polynomial models.
Similarly, the number of gate eigenvalues is exactly described by a quadratic with integer coefficients, \(N(d)=144d^2-180d+54\).
Substituting these three quadratics into \cref{eq:aces-figure-of-merit,eq:aces-variance} yields functional forms that precisely describe the scaling of \(\mathcal{F}\) and \(\mathcal{\sqrt{V}}\) with \(d\), which are also depicted in \cref{fig:unrotated-merit-distance-scaling}.
For depolarising noise, we fit the normalised traces of the gate eigenvalue estimator covariance matrix, obtaining functional forms for \(\mathcal{F}\) and \(\mathcal{\sqrt{V}}\) with relative errors of less than \(10^{-9}\) for each data point.
By contrast, for log-normal Pauli noise, we directly and simultaneously fit the functional forms for \(\mathcal{F}\) and \(\mathcal{\sqrt{V}}\).
Although the data represent estimates, we nevertheless obtain relative errors of less than \(10^{-3}\) for each data point.

Importantly, the functional forms derived by substituting the quadratics into \cref{eq:aces-figure-of-merit,eq:aces-variance} entail that in the limit of large \(d\), the figure of merit \(\mathcal{F}\), or expected normalised RMS error, approaches a constant value.
Also, the optimised experimental design requires \(219\) experiments, before Pauli frame randomisation, to estimate all of the circuit eigenvalues at all tested code distances \(d\), which range from \(d=3\) to \(d=17\).
This demonstrates that ACES is capable of estimating noise in surface code syndrome extraction circuits to a precision that is asymptotically independent of the number of qubits \(n\) in the code, using a number of experiments that is also asymptotically independent of \(n\).

\begin{figure}[b]
    \centering
    \includegraphics[width=\columnwidth]{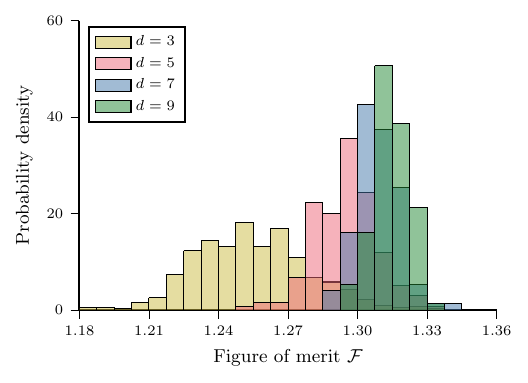}
    \vspace{-2em}
    \caption{Similar results as in \cref{fig:log-normal-merit-distribution}, but for the unrotated surface code.
    The histograms indicate the distribution of the figure of merit \(\mathcal{F}\), the expected normalised RMS error, of the optimised experimental design for the syndrome extraction circuit, across random instances of log-normal Pauli noise.
    Results are shown across a range of unrotated surface code distances \(d\).}
    \label{fig:unrotated-log-normal-merit-distribution}
\end{figure}

\begin{figure*}[t]
    \centering
    \includegraphics[width=\textwidth]{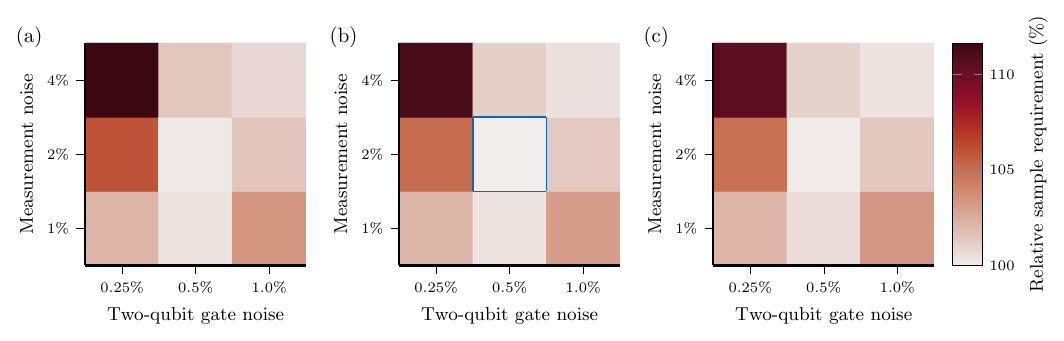}
    \vspace{-2em}
    \caption{Similar results as in \cref{fig:merit-heatmap}, but for the unrotated surface code.
    The experimental designs were optimised for depolarising noise with a range of error rates, with single-qubit gate error rates (a) \(r_1=0.0375\%\), (b) \(r_1=0.075\%\), and (c) \(r_1=0.15\%\), and two-qubit gate and measurement error rates indicated in the heatmaps.
    Experimental designs were evaluated for the syndrome extraction circuit of a distance-\(3\) unrotated surface code against \(400\) random instances of log-normal Pauli noise, whose average error rates are indicated by the blue square.
    Heatmap colour indicates the expected number of samples required to achieve a fixed estimation accuracy, expressed relative to the best-performing design optimised at the correct average error rates.
    Only three experimental designs require over \(10\%\) more samples to achieve the same accuracy as the best-performing design.}
    \label{fig:unrotated-merit-heatmap}
\end{figure*}

Notably, the optimised experimental design performs better on average when characterising log-normal Pauli noise, despite being optimised for depolarising noise.
Moreover, \cref{fig:unrotated-log-normal-merit-distribution} shows histograms of the figure of merit \(\mathcal{F}\) across random instances of log-normal Pauli noise for the syndrome extraction circuit at a range of unrotated surface code distances \(d\).
It shows that random samples of log-normal Pauli noise have increasingly similar figures of merit as \(d\) increases.
Together, these suggest that the performance of the optimised experimental design is not substantially harmed by small changes in the noise model that preserve average error rates.
The effects of randomness in the noise model on the performance of ACES also appear to average out with increasing system size.

We can also examine experimental designs optimised at different error rates to determine the robustness of their performance.
Specifically, we optimise for half the error rate, the error rate, and double the error rate, on single-qubit gates, two-qubit gates, and measurements, yielding \(27\) different experimental designs.
\Cref{fig:unrotated-merit-heatmap} shows the expected number of samples required to achieve a fixed estimation accuracy, evaluated over \(400\) random instances of log-normal Pauli noise for the syndrome extraction circuit of a distance-\(3\) unrotated surface code.
These values are expressed relative to the original experimental design optimised for the appropriate average error rates, which performs best.

Only \(5\) of the \(27\) experimental designs require over \(5\%\) more samples than the most performant experimental design to achieve the same expected accuracy, with the worst requiring \(11.64\pm 0.10\%\).
This corresponds to having a mean figure of merit of \(1.3236\pm 0.0014\), whereas the mean figure of merit of the most performant design is \(1.2527\pm 0.0012\).
The expected relative number of samples required to achieve the same accuracy is the square of this ratio, and is estimated precisely due to substantial covariance in the figure of merit between designs for each instance of log-normal Pauli noise.
Overall, this demonstrates that the performance of optimised experimental designs is robust to being optimised for inappropriate average error rates.

\begin{figure}[b]
    \centering
    \includegraphics[width=\columnwidth]{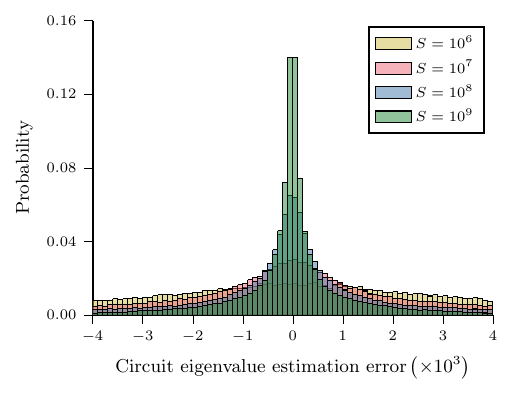}
    \vspace{-2em}
    \caption{Similar results as in \cref{fig:aces-eigenvalue-histogram}, but for the syndrome extraction circuit of a distance-\(17\) unrotated surface code.
    The histograms indicate distributions of the circuit eigenvalue estimation error (the difference between the estimated and true circuit eigenvalues).
    The data are for a fixed-seed random instance of log-normal Pauli noise over a range of measurement budgets \(S\).}
    \label{fig:unrotated-aces-eigenvalue-histogram}
\end{figure}

Finally, we can use our optimised experimental design to characterise a fixed-seed random instance of log-normal Pauli noise on the syndrome extraction circuit of a distance \(d=17\) unrotated surface code with \(n=1089\) qubits.
The optimised experimental design features \(25\) tuples which together estimate \(195,723\) circuit eigenvalues over \(219\) experiments, before Pauli frame randomisation, in order to estimate the \(38,610\) gate eigenvalues.
Optimising the experimental design at \(d=3\), generating it at \(d=17\), and then simulating ACES noise characterisation experiments for measurement budgets \(S\in\{10^6,10^7,10^8,10^9\}\) together took under \(3\) hours, with roughly half of that time being dedicated to stabiliser circuit simulations with Stim.
The fits in \cref{fig:unrotated-merit-distance-scaling} predict an average figure of merit \(1.3249\) and a standard deviation \(0.010\), across random instances of log-normal Pauli noise.
By comparison, for the simulated noise characterisation of a specific random instance of log-normal Pauli noise, the normalised RMS errors were \(\{1.3135, 1.3207, 1.3237, 1.3181\}\) across measurement budgets.
These results are consistent with our performance predictions.

\begin{figure*}[t]
    \centering
    \includegraphics[width=\textwidth]{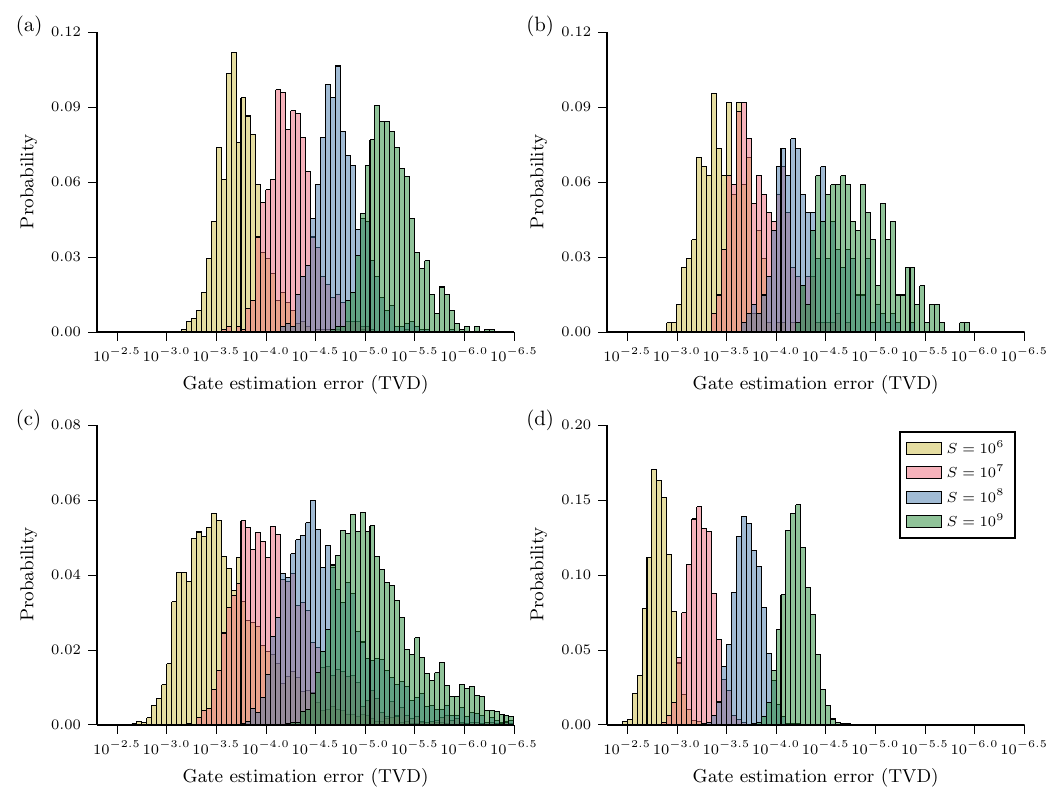}
    \vspace{-2em}
    \caption{Similar results as in \cref{fig:aces-tvd-histograms}, but for the syndrome extraction circuit of a distance-\(17\) unrotated surface code.
    The histograms indicate distributions of the gate estimation error (the total variation distance (TVD) between the estimated and true error probability distribution for the gate).
    The data are for a fixed-seed random instance of log-normal Pauli noise over a range of measurement budgets \(S\).
    The estimation error distributions are shown across the range of gate types appearing in the circuit: (a) dynamical decoupling \(X\) gates and padded identity gates, or Pauli gates; (b) Hadamard gates; (c) measurements; and (d) controlled-\(Z\) gates.}
    \label{fig:unrotated-aces-tvd-histograms}
\end{figure*}

The noise estimation procedure begins by estimating the circuit eigenvalues from raw measurement data.
\Cref{fig:unrotated-aces-eigenvalue-histogram} depicts histograms of the circuit eigenvalue estimation error, the difference between the estimated and true circuit eigenvalues, across the range of measurement budgets.
As the measurement budget increases, the distributions of the circuit eigenvalue estimation error narrow, straightforwardly improving the accuracy of the circuit eigenvalue estimates.

Ultimately, the protocol estimates the Pauli error probability distribution of all the gates and measurements appearing in the circuit.
We measure the gate estimation error with the total variation distance (TVD) between the estimated and true Pauli error probability distribution, a principled measure for probability distributions and half the diamond norm between Pauli channels~\citep{fawzi_lower_2023}.
\Cref{fig:unrotated-aces-tvd-histograms} depicts histograms of the gate estimation error across measurement budgets and the different types of gate appearing in the circuit, namely: padded identity gates, which are Pauli gates; Hadamard gates; measurements; and controlled-\(X\) gates.
The Pauli gates tend to be estimated to a higher precision than the Hadamard gates, despite both being single qubit gates.
By contrast, the two-qubit controlled-\(X\) gates are least accurately estimated, accounted for in part by their Pauli error probability distribution being over \(16\) errors, compared to \(4\) for the single-qubit gates and just \(2\) for measurements.

The gate estimation errors improve roughly by a constant factor of \(\sqrt{10}\) for each factor of \(10\) increase in the measurement budget \(S\), demonstrating the expected \(1/\sqrt{S}\) sample efficiency.
This is clear visually and quantified in \cref{tab:unrotated-median-tvd}, which shows the median gate estimation error across gate types and measurement budgets.
The Hadamard and controlled-\(X\) gate estimates outperform the trend, particularly for the smallest measurement budget \(S=10^6\).
This is because large estimation errors for small values of \(S\) are improved by ensuring parameter estimates are within bounds.
Specifically, gate eigenvalues greater than \(1\) are set to \(1\), and estimated probability distributions are projected into the simplex.
For the data in \cref{tab:unrotated-median-tvd}, at \(S=10^6\), \(460\) gate eigenvalues are set to \(1\), while at \(S=10^7\), only \(17\) gate eigenvalues are set to \(1\).

Overall, we see that the results for the rotated and unrotated surface code syndrome extraction circuits are very similar, supporting the utility of our methods for characterising noise in the syndrome extraction circuits of general topological quantum codes, as well as other fault-tolerant gadgets.

\begin{table}[b]
    \centering
    \vspace{-1.25em}
    \caption{Similar results as in \cref{tab:median-tvd}, but instead showing the medians of the gate estimation error distributions in \cref{fig:unrotated-aces-tvd-histograms}.}
    \vspace{0.25em}
    \def\arraystretch{1.4}
    \begin{tabular}{|c|c|c|c|c|}
        \hline
        \multirow[c]{2}{5.7em}{\centering Measurement budget} & \multicolumn{4}{c|}{\(-\log_{10}\) median gate estimation error (TVD)} \\
        \cline{2-5}
                 & Pauli & Hadamard & Measurement & Controlled-\(X\) \\
        \hline
        \(10^6\) & 3.728 & 3.466    & 3.538       & 2.826            \\
        \(10^7\) & 4.245 & 3.808    & 4.083       & 3.238            \\
        \(10^8\) & 4.725 & 4.290    & 4.585       & 3.712            \\
        \(10^9\) & 5.241 & 4.776    & 5.071       & 4.207            \\
        \hline
    \end{tabular}
    \label{tab:unrotated-median-tvd}
\end{table}

\end{document}